\documentclass[aps,showpacs,reprint,pre,numerical,showkeys]{revtex4-1}
\usepackage{amsmath,color,hyperref,amsfonts,graphicx,amssymb,esint}
\newcommand{\nd}{\mathrm{nd}}
\newcommand{\cotanh}{\mathrm{cotanh}}
\newcommand{\sn}{\mathrm{sn}}
\newcommand{\beq}{\begin{equation}}
\newcommand{\eeq}{\end{equation}}
\newcommand{\bea}{\begin{eqnarray}}
\newcommand{\eea}{\end{eqnarray}}
\newcommand{\On}{\mathcal{O}(\mathfrak{n})}
\newcommand{\nn}{\mathfrak{n}}
\newcommand{\dd}{\mathrm{d}}

\definecolor{rouge}{rgb}{0.84,0.18,0.07}
\definecolor{bleu}{rgb}{0.22,0.41,0.74}
\definecolor{vertf}{rgb}{0.08,0.46,0.07}
\hypersetup{colorlinks,urlcolor=vertf,citecolor=rouge,linkcolor=bleu,filecolor=black}

\begin{document}
\title{Resolvent methods for steady premixed flame shapes governed by the Zhdanov-Trubnikov equation}

\author{Ga\"{e}tan Borot}
\affiliation{Section de Math\'ematiques, Universit\'e de Gen\`eve, 2-4 rue du Li\`evre, 1206 Gen\`eve 4, Switzerland.}
\email{gaetan.borot@unige.ch}
\author{Bruno Denet}
\affiliation{Universit\'e d'Aix-Marseille, IRPHE, UMR 7342 CNRS, Technopole de Ch\^{a}teau-Gombert, 49 rue Joliot-Curie, 13384 Marseille Cedex 13, France.}
\email{bruno.denet@irphe.univ-mrs.fr}
\author{Guy Joulin}
\affiliation{Institut P', UPR 3346 CNRS, ENSMA, Universit\'e de Poitiers, 1 rue Cl\'ement Ader, BP 40109, 86961 Futuroscope Cedex, Poitiers, France.}
\email{guy.joulin@lcd.ensma.fr}

\begin{abstract}
Using pole decompositions as starting points, the one parameter ($-1 \leq c < 1$) nonlocal and nonlinear Zhdanov-Trubnikov (ZT) equation for the steady shapes of premixed gaseous flames is studied in the large-wrinkle limit. The singular integral equations for pole densities are closely related to those satisfied by the spectral density in the so-called $\On$ matrix model, with $\nn = -2\,\frac{1 + c}{1 - c}$. They can be solved via the introduction of complex resolvents and the use of complex analysis. We retrieve results obtained recently for $-1 \leq c \leq 0$, and we explain and cure their pathologies when they are continued naively to $0 < c < 1$. Moreover, for any $-1 \leq c < 1$, we derive closed-form expressions for the shapes of steady isolated flame crests, and then bicoalesced periodic fronts. These theoretical results fully agree with numerical resolutions. Open problems are evoked.
\end{abstract}


\maketitle

\section{Introduction}
\label{sec1}

Most flames propagating into premixed gaseous reactants may be viewed as fronts: their actual thickness $\ell \sim \sqrt{D_{\mathrm{th}}t_{\mathrm{th}}} \sim u_Lt_{\mathrm{ch}}$ (based on flat-flame speed $u_L$, heat diffusivity $D_{\mathrm{th}}$, and chemical time $t_{\mathrm{ch}}$) is often much smaller than the wavelength and amplitude of their deformations. Being also very subsonic (i.e. $u_L \ll$ speed of sound), such combustion fronts border fluids of constant and uniform densities: $\rho_u$ on the fresh side, and $\rho_b < \rho_u$ in the burnt gas. As the Atwood number $\mathcal{A} = \frac{\rho_u - \rho_b}{\rho_u + \rho_b} < 1$ is nonzero, such flames are subject to the hydrodynamic, hence nonlocal, Darrieus \cite{Darrieus}-Landau \cite{Landau} (DL) wrinkling instability at large wavelengths. At shorter scales though, the variations in local normal burning speed $u_n$ (relative to reactants) with front mean curvature \cite{Markstein} bring about a neutral wavelength $L_{\mathrm{neutral}}$ proportional to $\ell$, yet much longer in practice \cite{PelceClavin82}. Viewing wrinkled flames as fronts embedded in incompressible flows largely facilitates numerical simulations \cite{CretaMatalon} and theoretical analyses thereof.

Acknowledging that	$\mathcal{A} \ll 1$	implies a weak DL instability -- and hence asymptotically small front slopes and slow evolutions -- Sivashinsky analysis \cite{Sivashinsky} provided the first systematic weakly-nonlinear description of the local amplitude of a wrinkling: a companion numerical work \cite{MichelsonSivashinsky} confirmed that the Michelson-Sivashinsky (MS) equation (Eq.~(\ref{11}) with $c = 0$) correctly captures the slow spontaneous dynamics of flat-on-average flames if $\mathcal{A} \rightarrow 0^+$. In physical situations, $\mathcal{A}$ ranges from $0.2-0.5$ for thermonuclear flames in Supernovae \cite{Niemeyer} to $0.65-0.85$ for chemical fronts. It is thus important to incorporate higher orders in $\mathcal{A}$. The first two subleading orders can be absorbed in a renormalization of coefficients of the MS equation \cite{ClavinSivash,KazakLiberman,KAK05}. At the third one  a new quadratic term in the equation, but the fourth order correction can again be absorbed in coefficients \cite{KAKetal}. A similar nonlinearity first appeared in derivations of an equation for the wrinkling amplitude that assumed $\mathcal{A} = O(1)$ and postulated a small flame slope \cite{ZT,Bychkov}.

If a single space coordinate is retained, the Zhdanov-Trubnikov (ZT) type of equation \cite{ZT} so obtained for slow spontaneous evolutions of front wrinkles reads as follows, in rescaled form:
\beq
\label{11}\phi_t + \frac{1}{2}\big(\phi_x^2 + c\mathcal{H}[\phi_x]^2\big) = \nu\phi_{xx} - \mathcal{H}[\phi_x],
\eeq
where $c$ is some function of $\mathcal{A}$, $\phi(t,x)$ is a dimensionless front deformation about a flat shape ($\phi \equiv 0$) ; the subscripts denote derivatives with respect to rescaled time $(t)$ or coordinate $(x)$. For space-\emph{periodic cells}, $0 < \nu < 1$ will denote the neutral-to-actual wavelength ratio, provided the reference length is suitably chosen. For the \emph{isolated crests} of infinite wavelengths, $\nu > 0$ could be rescaled to unity by a change of variables but it is kept in \eqref{11} for comparisons with cells. The dependence of $u_n - u_L$ on curvature gave the $\nu\phi_{xx}$ term in \eqref{11}, while the Hilbert transform:
\beq
\mathcal{H}[\phi_x] = \frac{1}{\pi}\fint \frac{\phi_{x}(x')\,\dd x'}{x - x'},
\eeq
in the right-hand side encodes the nonlocal DL instability: a small $\phi(t,x) \propto \exp(\varpi t + {\rm i}\kappa x)$ indeed has $\varpi = |\kappa| - \nu\kappa^2$ as growth rate. The stabilizing local nonlinearity in \eqref{11} combines geometry \cite{Sivashinsky} and hydrodynamics \cite{ClavinSivash}. The nonlocal one, absent from the MS equation ($c = 0$), is mainly fluid-mechanical since $\phi_x^2 + \mathcal{H}[\phi_x]^2$ is proportional to the kinetic energy of wrinkling-induced flow disturbances. It may be destabilizing if $c < 0$, or overstabilizing if $c > 0$, affecting the solutions of \eqref{11} in several regards, e.g. at the front tips (local, usually sharp maxima of $\phi$).

Besides an interest \textit{per se}, solving the nonlinear and nonlocal \eqref{11} analytically with $c$ as general as possible -- and viewed here as a free parameter -- might help one fit flame shapes from experiments \cite{Quinard_ftc2011} ; or from simulations that use $\mathcal{A} = O(1)$ \cite{CretaMatalon}. This could also form the basis of stability analyses of curved fronts, e.g. generalizing that of \cite{VaynblattMatalon} to large wrinkles. The task is facilitated by the fact that, just like the MS equation \cite{TFH85}, \eqref{11} where $x$ is replaced by a complex variable $Z = x + {\rm i}B$, admits meromorphic solutions with pairs of complex-conjugate simple poles $Z_{-k}(t) = Z_{k}^*(t)$. When restricted to the real line $Z = x$, these solutions are real-valued and represent the front slope. Then, \eqref{11} is converted into coupled nonlinear differential equations for the poles $Z_k(t)$ \cite{joulin91}. When the front acquires a steady shape, those poles of $\phi_x$ that remain at rest for $t \rightarrow \infty$ generically are aligned along parallels to the imaginary axis, due to the same mechanism as for $c = 0$ \cite{TFH85}: at $Z_k = {\rm i}B_k\,\,(\mathrm{mod}\,2\pi)$ when $x$-periodicity is assumed, and also at $Z_m = \pi + {\rm i}b_m\,\,(\mathrm{mod}\,2\pi)$ if Neumann boundary conditions are employed (or periodic ones with $x \leftrightarrow - x$ symmetry).
In any such steady configurations, the equations that govern the $2N$ motionless pole altitudes may be viewed as expressing an electrostatic (or kinematic) equilibrium under mutual pairwise repulsions between charges (or fluid sources), all subject to an external field originating from the DL instability mechanism. When $N\gg 1$, corresponding to large wrinkles, the spacing between consecutive poles scales like $B_N/N$, enabling one to adopt a continuous approximation of the equilibrium conditions, and rewrite them as \emph{integral equations for pole densities} \cite{TFH85}. As first shown for $c = 0$ in \cite{joulindenet08}, and confirmed when $c < 0$ \cite{joulindenet12}, a key step to describe the structure of large steady wrinkles is to first solve the simpler case of isolated crests ; these have infinite wavelengths and poles at $Z_k = \mathrm{i}B_k$ in arbitrary number $2N$.

In \cite{joulindenet12}, such integral equations were solved when $-1 \leq c \leq 0$ by exploiting peculiar properties of index-$\frac{1}{2}$ Meixner-Pollaczek orthogonal polynomials evidenced in \cite{dunkl}, firstly for isolated crests and then for monocoalesced periodic fronts (a single maximum of $\phi$ per wavelength). In both cases, the results nicely agreed with numerical resolutions of the discrete pole equations when $N \gg 1$. But the naive continuation to $c > 0$ led to unsatisfactory, yet intriguing results. For example, the front slope so obtained for isolated crests with $c > 0$ reads
\begin{equation}
\label{12}\phi_x(x) = -\frac{1}{\sqrt{c}}\sin\Big[\gamma\sinh^{-1}\Big(\frac{B_*}{x}\Big)\Big],
\end{equation}
with $B_{*} = \frac{4}{\gamma}\,\frac{\sqrt{c}}{1 - c}\,\nu N$ and $c = \tanh^2(\pi\gamma/2)$ ; this lies surprisingly close to the numerically-determined slope profile whenever $0 \leq c \lesssim 0.6$, except very close to $x = 0$ where \eqref{12} predicts an essential singularity. The pole density associated with \eqref{12} also stays very close to its numerical counterpart, except at $|B| \ll B_*$, where it oscillated infinitely many times and becomes negative at places: this is disallowed, for pole densities must be nonnegative. As conjectured in \cite{joulindenet12}, such localized pathologies might partly result from the failure of an assumption implicitly made in the continuation from $c \leq 0$ to $c > 0$, namely that the density support would still include $B = 0$, instead of being limited to $0 < B_{\min} \leq |B| \leq B_{\max} \approx B_{*}$ for some $B_{\min} > 0$, and one may imagine that $B_{\min}/B_{\max}$ increases rapidly with $c$. An analysis of the discrete pole equations for $c \rightarrow 1^{-}$ supports the first conjecture, and the results of Sec.~\ref{sec3}-\ref{sec4} will confirm and quantify both.

To this end, another method is devised to compute the flame shapes analytically. The key step is to introduce an appropriate \emph{resolvent}, which can be interpreted as a complex electric (or velocity) field caused by charges (or fluid sources) distributed according to the unknown pole density profile. The equations that rule the positions of the poles then turn into a nonlocal, scalar Riemann-Hilbert problem for the resolvent, which might be solved by complex analysis techniques. The advantage of this method is that, once the resolvent is found, the pole density and the front slope can be obtained easily, without extra integration.

In this way, we reproduce the results obtained in \cite{joulindenet12} for $-1 \leq c \leq 0$, and explain and fix the pathologies of \eqref{12} for $c > 0$. The method is valid whatever the sign of $c$, and generalizes results on wrinkled fronts with two unlike crests per wavelength, which was so far available for the MS equation only \cite{joulindenet08}. In all the configurations envisaged, we compare the analytical predictions with numerical determinations of the discrete $B_k$ and of $\phi$.

The paper is organized as follows. In Sec.~\ref{sec2}, the pole decompositions and their continuous approximations are introduced. We introduce at a formal level the notion of resolvent and we give equivalent forms of the ZT equation in \S~\ref{sec22}. The resolvent approach is then exploited for isolated crests (Sec.~\ref{sec3}), and next adapted to two-crested $x$-periodic fronts (Sec.~\ref{sec4}) ; in both cases comparisons with numerical results are presented and discussed. We summarize our results and formulate open problems concerning flame shapes in Section~\ref{sec5}. A resolution of the discrete pole equations for $c \rightarrow 1^{-}$, and technical details about resolvent determinations, are presented in Appendices~\ref{appA} and \ref{appC} respectively.

We also find links between flames fronts governed by the ZT equation, and other statistical-mechanical problems inspired from random matrix theory. Since the main body of the article is devoted to flames, we postpone this discussion to Appendix~\ref{appE}.

\section{Pole decompositions, isolated crests and the limit of large crests}
\label{sec2}

\subsection{Discrete poles}
\label{sec21}
Once properly shifted as to be centered, steady isolated crests of infinite wavelengths have
$\phi = \phi(x)$ where $\phi_x$ is a sum of $2N$ simple pole contributions $\propto \frac{1}{x - iB_k}$, $|k| = 1,\ldots,N$. $B_{-k} = -B_k$ ensures that $\phi(x)$ is real valued when $x$ is real. Since $\mathcal{H}[\frac{1}{x - {\rm i}B}] = \frac{-{\rm i}\mathrm{sgn}(B)}{x - {\rm i}B}$, the dominant balance in \eqref{11} implies that each pole has residue $\frac{-2\nu}{1 - c}$:
\beq
\label{21}\phi_x = \frac{-2\nu}{1 - c}\sum_{\substack{k = -N \\ k \neq 0}}^{N} \frac{1}{x - {\rm i}B_k}.
\eeq
For \eqref{11} to be satisfied, the pole altitudes $B_k$ obey coupled equations, originally expressed for $k =\pm 1,\ldots,\pm N$ as \cite{joulin91}:
\beq
\label{22}\sum_{\substack{j = -N \\ j \neq 0,k}}^{N} \frac{1 - c\,\mathrm{sgn}(B_k)\,\mathrm{sgn}(B_j)}{(1 - c)(B_k - B_j)} = \frac{\mathrm{sgn}(B_k)}{2\nu}.
\eeq
Without loss of information, \eqref{22} may be rewritten in terms of $B_1,\ldots,B_N > 0$ only. For $k = 1,\ldots,N$:
\beq
\label{23}\sum_{\substack{j = 1 \\ j \neq k}}^{N} \frac{1}{B_k - B_j} + \frac{1 + c}{1 - c}\sum_{j = 1}^N \frac{1}{B_k + B_j} = \frac{1}{2\nu},
\eeq
whereby the discontinuous $\mathrm{sgn}(B)$ function no longer shows up.

Summing $B_k \cdot$(Eq.~\eqref{23}) yields $0 < B_1 + \cdots + B_N = \nu N\big(\frac{2N}{1 - c} - 1)$, whence $c < 1$.  This must be considered jointly with the constraint $-1 \leq c$ needed from $B_1/\nu > 0$ (obtained for $c \rightarrow -1$, cf. \cite{joulindenet12}). For any fixed $-1 \leq c < 1$, the $B_k$'s will scale like $\nu N$ if $N \rightarrow \infty$, as in the $c = 0$ case, whereby the typical spacing $B_k - B_{k - 1}$ is $\sim \nu \sim B_N/N$. One may thus introduce a density
\beq
\label{positivity}\varrho(B) \geq 0
\eeq
here such that $\varrho(B)\dd B$ measures the number of poles with \emph{positive} altitudes in $[B,B+\dd B]$. This continuous approximation enables to rewrite \eqref{23} as an integral equation, valid for $B \in [B_{\min},B_{\max}]$:
\beq
\label{24}\fint_{B_{\min}}^{B_{\max}} \frac{\varrho(B')\,\dd B'}{B - B'} - \frac{\nn}{2}\int_{B_{\min}}^{B_{\max}} \frac{\varrho(B')\,\dd B'}{B + B'} = \frac{1}{2\nu},
\eeq
where $\fint$ denotes the Cauchy principal value, and the constant $\nn$ is defined as:
\beq
\label{nc}\nn = -2\,\frac{1 + c}{1 - c}.
\eeq
Notice that $-1 < c < 0$ is mapped to $-2 < \nn < 0$, and it will be convenient to set in this regime $c = -\tan^2(\pi\gamma/2)$, so that: 
\beq
\label{ncos} 0 < \gamma < 1/2,\qquad \nn = -2\cos(\pi\gamma).
\eeq
When $0 < c < 1$, we rather have $\nn < -2$, and a convenient parametrization is $c = \tanh^2(\pi\gamma/2)$, so that:
\beq
\label{ncosh}0 < \gamma,\qquad \nn = -2\cosh(\pi\gamma).
\eeq
The equation \eqref{24} must be solved subjected to the normalization condition:
\beq
\label{26}\int_{B_{\min}}^{B_{\max}} \varrho(B)\,\dd B = N,
\eeq
and the continuous version of \eqref{21} will express the front slope $\phi_x$ in terms of $\varrho(B)$ as:
\beq
\label{27}\phi_x(x) = \frac{4\nu}{1 - c}\,\mathrm{Im}\Big(\int_{B_{\min}}^{B_{\max}} \frac{\varrho(B)\,\dd B}{{\rm i}x - B}\Big).
\eeq
The above $\varrho(B)$ may be nonzero for $B \geq 0$ \emph{only}, contrary to the two-sided even density $\rho(B) = \varrho(B) + \varrho(-B)$ employed in \cite{joulindenet12}. Also notice that \eqref{24} allows for a lower end $B_{\min} \geq 0$ of the density support, besides the upper end $B_{\max}$ ; both have to be determined as parts of the solution. For $c \leq 0$, $B_{\min}$ happened to vanish \cite{joulindenet12}, whereas an argument presented in Appendix~\ref{appA} shows that $B_{\min} > 0$ in the limit $c \rightarrow 1^{-}$. One of the outcome of this article is that $B_{\min} > 0$ for any $0 < c < 1$ (see Fig.~\ref{Fi1}), while $B_{\min} = 0$ for any $-1 < c \leq 0$.

\subsection{Preliminary comments}

The kind of systems \eqref{22}, which express a repulsion between the poles $B_k$ and some of their images (here the $B_{-k} = -B_k)$ with possibly different intensity (here, if $k,k' \geq 1$, the interaction of $B_k$ and $B_{-k'}$ is $(-\nn/2)$ times stronger than the mutual interaction between $B_{k}$ and $B_{k'}$), is ubiquitous in statistical physics. In the context of quantum integrable system, they typically appear as the result of a \emph{Bethe Ansatz} to determine the eigenvalues of transfer matrices. This repulsion phenomenon is also observed for the zeros of orthogonal polynomials \cite{Marcellan}, optimal approximation points and the eigenvalues of random matrices \cite{Defcours,Forrester}. The connection between those topics and flame fronts is not accidental: we point out in Appendix~\ref{appE} that the ZT equation coincides with the large $N$ limit of a Schwinger-Dyson equation in a statistic-mechanical ensemble of $N$ particles called "$\beta$-deformed $\On$ matrix model".

The problem of describing how poles satisfying an equation like \eqref{22} condense when $N$ is large, i.e. of solving singular integral equations like \eqref{24}, is sometimes more convenient to consider from the point of view of complex analysis. There often exists an appropriate Riemann surface containing the support of $\varrho$, on which the density can be continued to an analytic function $\tilde{\varrho}$. One can then use the powerful tools of algebraic geometry to determine it, in particular the behavior of $\tilde{\varrho}$ at its singularities fixes much of the solution. We may bring up in the same spirit the techniques of conformal mapping used to solve Dirichlet problems for harmonic functions, which is well-known in electrostatics. 
The techniques to solve \eqref{24} effectively have been developed prior to this work in the context of matrix models, for all values of $c$ \cite{GaudinKostov,Kostov,EynardKristjansen,EynardKristjansen2}. In this article, we adapt them to the determination of flame shapes.

\begin{figure} 
\includegraphics[width=8cm]{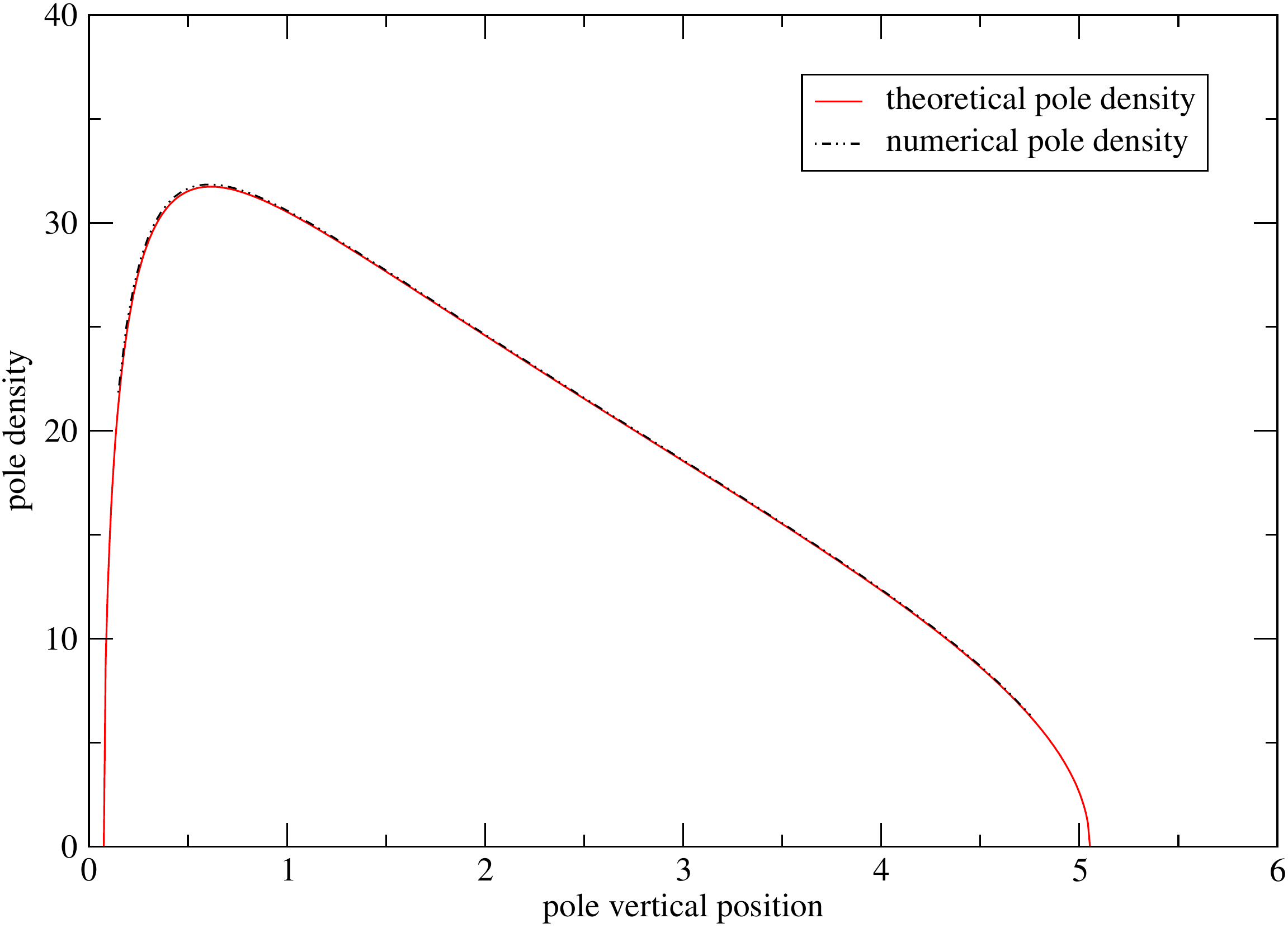}
\includegraphics[width=8cm]{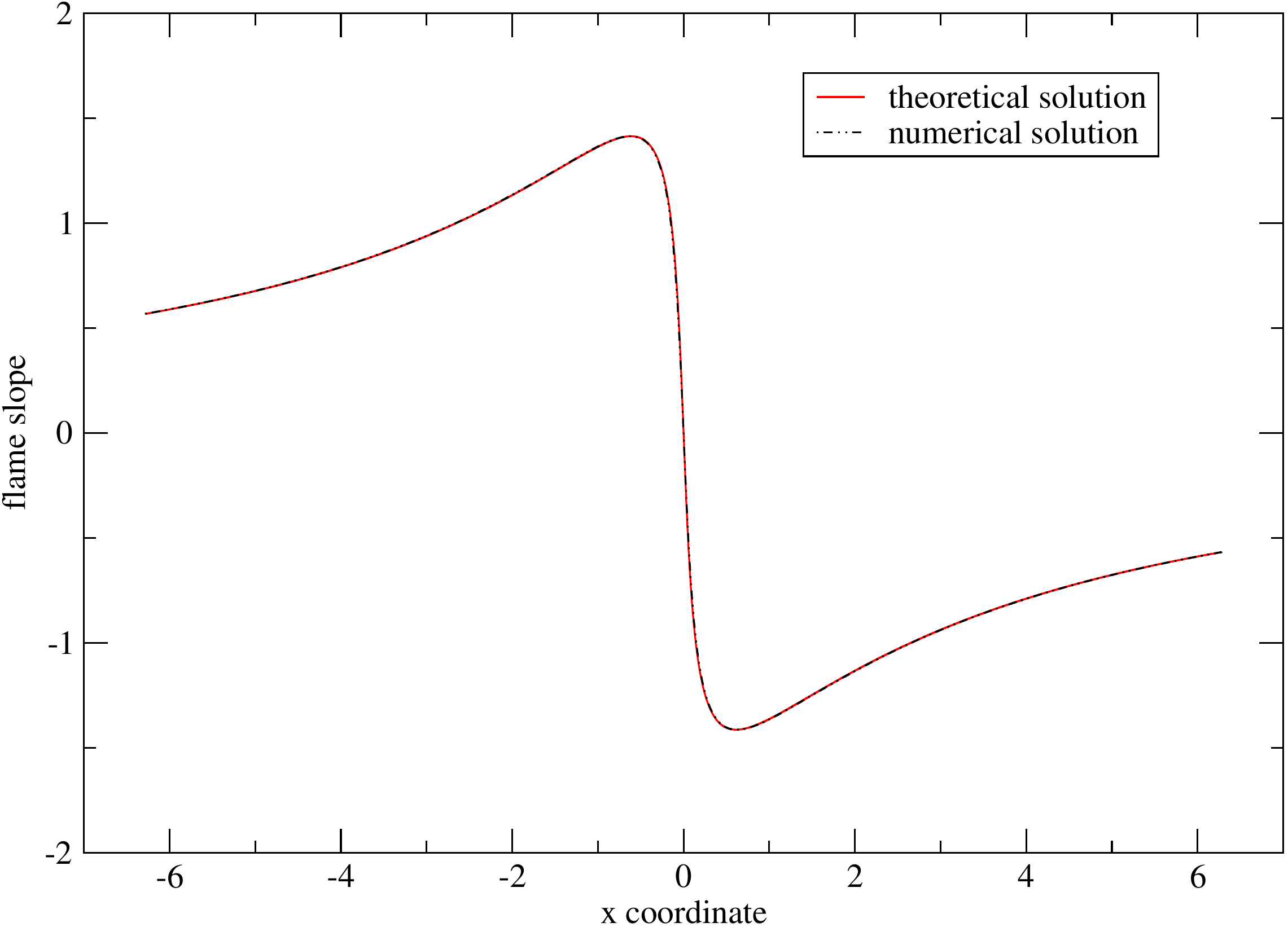}
\caption{\label{Fi1} Comparison of theoretical and numerical pole density (top, theoretically given by \eqref{35}) and flame slope (bottom, theoretically given by \eqref{sll}) for an isolated crest with $c > 0$. The plot assumes $c = 0.5$, $N = 100$ and $1/\nu = 199.5$.}
\end{figure}

Before going into details, it is noteworthy that the behavior of $\varrho$ at the edges of the supports features some universality: it depends only on the type of short-distance repulsion between poles, and does not on the details of external potential (the right-hand side in \eqref{24}). If $B_k$ remains away from their images, the density will typically behave as a squareroot near $B_{\min} > 0$. With \eqref{24} this happens for $c > 0$, and obviously the flame slope $\phi_x(x)$ will be smooth near $x = 0$. But if $B_{\min} = 0$, the poles close to $0$ are subjected to a strong self-interaction but also a strong interaction with their images at opposite position since the latter are also close to $0$. The exponent $\alpha$ such that $\varrho(B) \propto B^{-\alpha}$ when $B \rightarrow 0^{+}$ now depends continuously on the relative strength of interactions (see Sec~\ref{sec31})
\beq
\label{expo}\alpha = \gamma = \frac{1}{\pi}\cos^{-1}(-\nn/2) = \frac{2}{\pi}\tan^{-1}(\sqrt{-c})
\eeq
With \eqref{24}, this results for $-1 < c \leq 0$ in a cusp of the flame profile at $x = 0$, such that $\phi_x(x) \propto -\mathrm{sgn}(x)|x|^{-\gamma}$.

Anticipating on Sec.~\ref{sec3}, we find that the exact solution of \eqref{24} nicely agrees with what the numerical resolution of the discrete pole equations \eqref{23} gives for $N \gg 1$, and with the slope $\phi_x(x)$ ensuing from \eqref{27}. In the regime $-1 < c \leq 0$, this exact solution is constructed in \S~\ref{aisa} and was already known and compared to numerics in \cite[Fig. 3]{joulindenet12} ;  the corresponding flame profile has a cusp as we just discussed. For $0 < c < 1$, the exact solution is devised in \S~\ref{sec31} and displayed in Fig.~\ref{Fi1}. Unlike \cite[Fig. 5]{joulindenet12} where the pathological formula \eqref{12} was plotted, the theoretical prediction in Fig.~\ref{Fi1} does not feature a singularity at $x = 0$, as it should be (we used in both plots the exact same parameters).

\subsection{Continuum of poles and resolvents}
\label{sec22}
A function $\phi_x(x)$ (assumed square-integrable) can always be represented uniquely as \cite{Titchmarsh}:
\beq
\label{Wix}\phi_x(x) = \frac{4\nu}{1 - c}\,\mathrm{Im}\,W({\rm i}x)
\eeq
where $W(z)$ is a holomorphic function in the domain $\{\mathrm{Re}\,z < 0\}$, called \emph{resolvent}.
In this paragraph -- which can be read independently -- we clarify what does it mean for $\phi_x(x)$ to solve the ZT equation in terms of the resolvent, and therefore put the discrete pole decompositions \eqref{21} in a broader perspective. A more elementary approach will be adopted in \S~\ref{extr}.

Let us look for solutions of ZT such that $W(z)$ is actually holomorphic in $\mathbb{C}\setminus\mathcal{C}$, where $\mathcal{C}$ is a reunion of bounded arcs in the region $\{\mathrm{Re}\,z \geq 0\}$, which is invariant by complex conjugation. We allow $\mathcal{C}$ to touch finitely many times the axis ${\rm i}\mathbb{R}$: these points of contact would correspond to singularities of $\phi_x(x)$. Besides, we assume that the front is asymptotically flat, i.e. $W(z) = o(1)$ when $z \rightarrow \infty$. Then, we can use Cauchy residue formula to represent:
\beq
W(z) = \int_{\mathcal{C}} \frac{\varrho(B)\,\dd B}{z - B}
\eeq
where we introduced the density:
\beq
\varrho(B) = \frac{W(B - {\rm i}0) - W(B - {\rm i}0)}{2{\rm i}\pi}
\eeq
The discrete pole decomposition \eqref{21} corresponds to the special case where $\varrho(B)$ is a sum of $N$ Dirac masses located at $B = B_k$ for $k = 1,\ldots,N$. Here, we rather want to consider the case where $\varrho(B)$ is continuous. 

The assumption $\mathcal{C} = \mathcal{C}^*$ allows to rewrite \eqref{Wix} as:
\beq
\phi_x(x) = \frac{2\nu}{{\rm i}(1 - c)}\big[W({\rm i}x) - W(-{\rm i}x)\big],
\eeq
when $x \in \mathbb{R}$ is not at a singularity of $\phi_x(x)$. Thanks to the assumed analytic properties of $W(z)$, the Hilbert transform of $\phi_x(x)$ can be computed explicitly:
\beq
\mathcal{H}[\phi_x](x) = \frac{2\nu}{1 - c}\big[W({\rm i}x) + W(-{\rm i}x)\big].
\eeq
Therefore, the ZT equation is equivalent to: for any $z \in {\rm i}\mathbb{R}$,
\bea
\label{ZTW}& & W^2(z) + W^2(-z) + \nn W(z)W(-z)  \\
& & + W'(z) + W'(-z) - (1/\nu)\big[W(z) + W(-z)\big] =  0. \nonumber
\eea
where $\nn$ was defined in \eqref{nc}. And, since \eqref{ZTW} is an equality between analytic functions, it must be valid in the whole domain of analyticity, namely for any $z \in \mathbb{C}\setminus\mathcal{C}$.

Large wrinkles correspond to the limit $\nu \rightarrow 0$ but with $W(z)$ scaling like $1/\nu$ so as to keep a macroscopic front slope in \eqref{Wix}. In this limit, the derivative term in \eqref{ZTW} can be neglected and we find $\mathcal{E}[W](z) = 0$, where we set:
\bea
\label{ZZZ}\mathcal{E}[W](z) & = & W^2(z) + W^2(-z) + \nn W(z)W(-z) \nonumber \\
& & - (1/\nu)\big[W(z) + W(-z)\big].
\eea
Computing the discontinuity of $\mathcal{E}[W](z)$ at a point $B \in \mathcal{C}$, we find:
\bea
& & [W(B + {\rm i}0) - W(B - {\rm i}0)] \\
& & \times [W(B + {\rm i}0) + W(B - {\rm i}0) + \nn W(-B) - 1/\nu] = 0. \nonumber
\eea
Therefore:
\beq
\label{dsicc}W(B + {\rm i}0) + W(B - {\rm i}0) + \nn W(-B) = 1/\nu.
\eeq
In terms of the density $\varrho(B)$, this equation is equivalent to \eqref{24} which has been derived in \S~\ref{sec21} for the continuum limit of a discrete pole configuration.

Conversely, imagine that we have a function $W(z)$ is holomorphic in a domain $\mathbb{C}\setminus\mathcal{C}$, which decays at infinity, which has a continuous density $\varrho(B)$, and which satisfies \eqref{dsicc}. It follows that $\mathcal{E}[W](z)$ is holomorphic on $\mathbb{C}\setminus(\mathcal{C}\cup -\mathcal{C})$ and decays at infinity. Furthermore, by multiplying \eqref{dsicc} by $W(B + {\rm i}0) - W(B - {\rm i}0)$ and rearranging, we find that $\mathcal{E}[W](z)$ is continuous across $\mathcal{C}$. By parity, it is also continuous across $-\mathcal{C}$, and therefore, $\mathcal{E}[W](z)$ is an entire function decaying at infinity. By Liouville theorem, it must vanish: in other words, $\phi_x(x)$ given by \eqref{Wix} is solution to the ZT equation \eqref{11} in the limit of large wrinkles.

\section{Resolvents and isolated crests}
\label{sec3}

\subsection{Strategy}
\label{extr}

Even without refering to \S~\ref{sec22}, the form of \eqref{24} suggests to introduce the so-called resolvent, defined for a complex variable $z$ as:
\beq
\label{31}W(z) = \int_{B_{\min}}^{B_{\max}} \frac{\varrho(B)\,\dd B}{z - B}.
\eeq
In the context of an electrostatic (of a fluid-mechanical) analogy, $W(z)$ would represent the complex electric (or velocity) field generated by charges (or fluid sources) deposited according to the density $\varrho(B)$ along the segment $[B_{\min},B_{\max}]$ in the $z$-complex plane. Equivalently, $W(z)$ is the unique function which is holomorphic on $\mathbb{C}\setminus[B_{\min},B_{\max}]$, behaves as $N/z$ when $z \rightarrow \infty$, and is discontinuous on $[B_{\min},B_{\max}]$ with a given jump:
\beq
\label{32}\frac{W(B - {\rm i}0) - W(B + {\rm i}0)}{2{\rm i}\pi} = \varrho(B).
\eeq
The Cauchy principal value in \eqref{24} can then be expressed as:
\beq
\label{333}\fint_{B_{\min}}^{B_{\max}} \frac{\varrho(B')\,\dd B'}{B - B'} = \frac{W(B - {\rm i}0) + W(B + {\rm i}0)}{2}.
\eeq
We can also reformulate \eqref{24} in terms of the resolvent only: for $B \in [B_{\min},B_{\max}]$,
\beq
\label{34}W(B + {\rm i}0) + W(B - {\rm i}0) + \mathfrak{n}W(-B) = 1/\nu.
\eeq
The normalization condition \eqref{26} is translated into the requirement:
\beq
\label{infinity}W(z) \sim N/z,\qquad z \rightarrow \infty,
\eeq
and the front slope can be expressed readily by comparing of \eqref{27} and \eqref{31}:
\beq
\label{slope}\phi_x(x) = \frac{4\nu}{1 - c}\,\mathrm{Im}\big[W({\rm i}x)\big].
\eeq

The strategy to solve \eqref{34} has been developed in \cite{EynardKristjansen,EynardKristjansen2}. The first step is to assume that the position of the density support $[B_{\min},B_{\max}]$ is given, and determine it by consistency only at the end. The second step consists in performing a change of variable which uniformizes the complex plane with two cuts $[B_{\min},B_{\max}]$ and $[-B_{\max},-B_{\min}]$, since points on these two segments are involved in \eqref{34}. In other words, one constructs:
\beq
\omega(\psi) = W(z(\psi)),
\eeq
where $\psi$ belongs to some domain $\mathcal{D}$ in the complex plane, so that \eqref{34} relates boundary values of $\omega$ on $\mathcal{D}$, in such a way that its resolution becomes "easy" using Schwarz reflection principle. This can be done with trigonometric functions if $B_{\min} = 0$, whereas elliptic functions shows up if $B_{\min} > 0$. The third step is the resolution of the equation for $\omega$ taking into account its analytical properties and the normalization \eqref{26}, which result in the determination the position of the support $[B_{\min},B_{\max}]$. If several solutions for the support are available, there is in general a unique one which ensures positivity of the density \eqref{positivity}. We already announce the result that $B_{\min} = 0$ whenever $-1 < c < 0$, while $B_{\min} > 0$ in the case $0 < c < 1$, and we now explain the implementation of this method in both cases. The limit cases $c = -1,0,1$ are better discussed separately, respectively in \cite{joulindenet12}, \cite{joulindenet08} and Appendix~\ref{appA}.

\subsection{Solution for $0 < c < 1$}
\label{sec31}
\begin{figure}
\includegraphics[width=8.7cm]{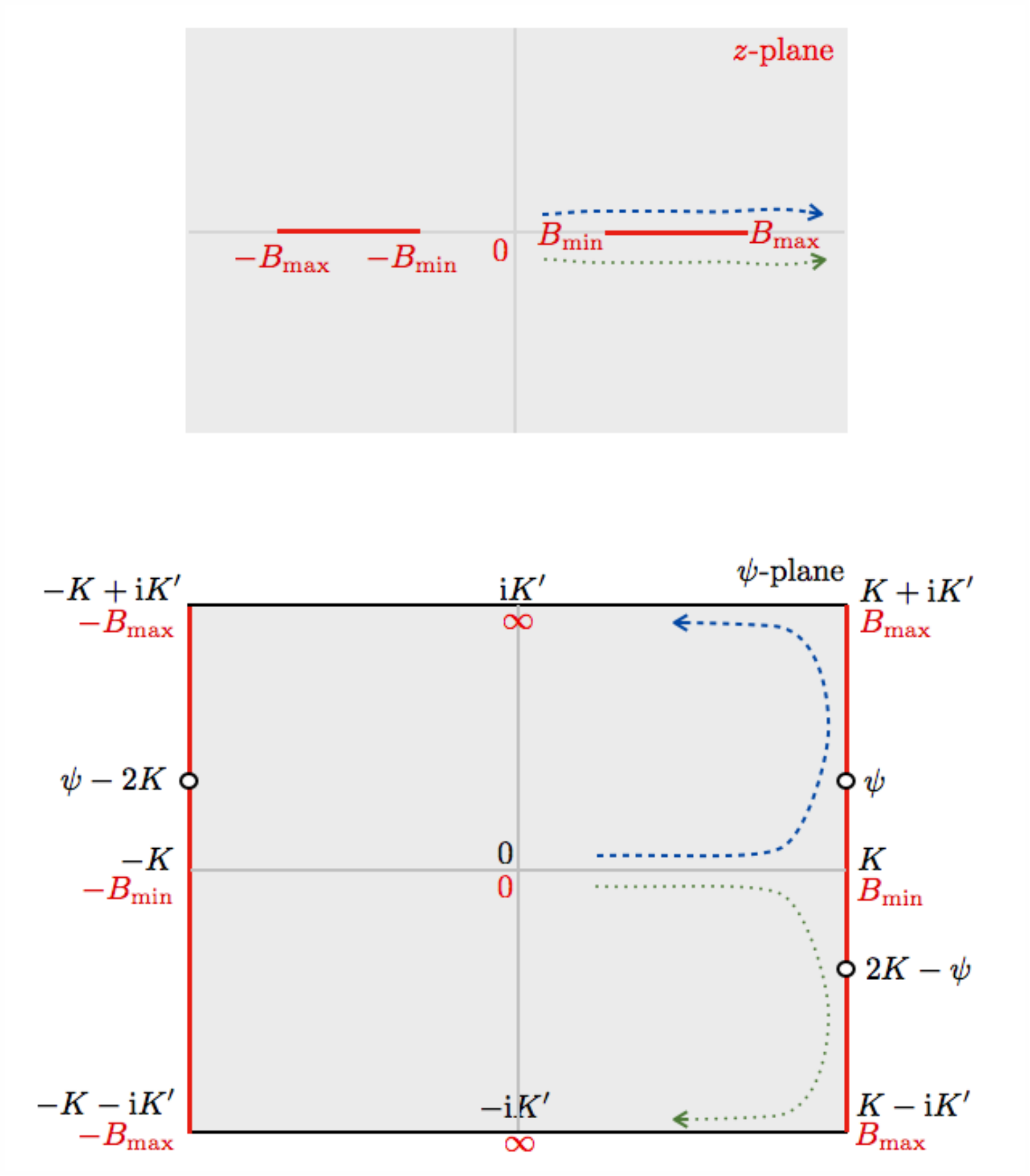}
\caption{\label{mapp} Mapping between the $z$-plane (top panel) with its two cuts on $[-B_{\min},-B_{\max}]\cup[B_{\min},B_{\max}]$ and the rectangle $\mathcal{D}$ in the $\psi$-plane (lower panel). For special points in the $\psi$-plane, we indicate below (in red) the corresponding value of $z$.
The dashed blue and dotted green paths indicates how the mapping "opens" the cuts.}
\end{figure}

We assume in this paragraph $B_{\min} > 0$, and later find this implies $0 < c < 1$. It is convenient to choose as change of variable:
\beq
\label{paramtrigo}z = B_{\min}\,\sn_k(\psi),\qquad k = \frac{B_{\min}}{B_{\max}},
\eeq
where $\sn$ is the Jacobi elliptic sine function \cite{GuoWang}. Let $K$ and $K'$ be the complete elliptic integrals with modulus $k$. The $z$-complex plane with its two cuts $[-B_{\max},-B_{\min}]\cup[B_{\min},B_{\max}]$ is mapped onto the rectangle $\mathcal{D}$ of vertices $\pm K$ and $\pm {\rm i}K'$ (see Fig.~\ref{mapp}). More precisely: the segment $[B_{\min},B_{\max}] \pm {\rm i}0$ is mapped to the half-edge $[K,K \pm {\rm i}K']$ in the rectangle ; the lower (resp. upper) side of the segment $]-\infty,-B_{\max}]\cup[B_{\max},+\infty[$ is mapped to the upper (resp. lower) horizontal edge of the rectangle, and in particular the point $z = \infty$ correspond to $\psi = \pm{\rm i}K'$ ; and the segment $]-B_{\max},-B_{\min}[ \pm {\rm i}0$ is mapped to $[-K,-K\pm {\rm i}K']$.

$\omega(\psi) = W(z(\psi))$ defines a holomorphic function in the rectangle $\mathcal{D}$. The analytical properties of the resolvent \eqref{31} provide relations between values of $\omega$ on the boundary of $\mathcal{D}$. Firstly, since the resolvent is continuous across $]-\infty,-B_{\max}]\cup[B_{\max}+\infty[$,  we must have:
\beq
\label{35a}\omega(\psi) = \omega(\psi + 2{\rm i}K'),
\eeq
when $\psi$ belongs to the lower horizontal edges of $\mathcal{D}$. Similarly, the continuity of $W(z)$ across $[-B_{\max},-B_{\min}]$ implies:
\beq
\label{36a}\omega(\psi) = \omega(-2K - \psi),
\eeq
when $\psi$ belongs to the edge $[-K - {\rm i}K',- K + {\rm i}K']$. Eventually, \eqref{34} itself is equivalent to:
\beq
\label{37}\omega(\psi) + \omega(\psi - 4K) + \nn\,\omega(\psi - 2K) = 1/\nu,
\eeq 
when $\psi$ belongs to the edge $[K - {\rm i}K',K + {\rm i}K']$.

The key observation is that successive applications of Schwarz reflection principle and \eqref{35a}-\eqref{37} allows to continue analytically $\omega$ on the whole $\psi$-complex plane, so that \eqref{35a}-\eqref{37} are satisfied \emph{everywhere}. Indeed, \eqref{35a} allows to define $\omega(\psi)$ as a $2{\rm i}K'$-periodic function on the strip $\mathcal{S} = \{-K < {\rm Re}\,\psi < K\}$, while \eqref{36a} allows to define $\omega(\psi)$ on the doubled strip $\mathcal{S}_2 = \{-3K < {\rm Re}\,\psi < K\}$. Then, \eqref{37} provides a relation between values of $\omega$ on the boundary of $\mathcal{S}$. Furthermore, if we take a point $\psi \in \mathcal{S}_2 - 4K$, the points $\psi - 2K$ and $\psi$ belong to $\mathcal{S}_2$. So, enforcing \eqref{37} defines $\omega(\psi)$ as an analytic function of $\psi \in \mathcal{S}_2' = \mathcal{S}_2 \cup(\mathcal{S}_2 - 4K)$, which satisfies \eqref{37} everywhere on $\mathcal{S}_2'$. This argument can be used recursively to analytically continue $\omega(\psi)$ on all strips of the form $\mathcal{S}_2 + 4mK$ for any integer $m$, so as to cover the whole complex plane.

The problem is now reduced to an easier one, namely finding the most general analytic functions which satisfy \eqref{35a}-\eqref{37}, i.e. with an appropriate behavior under $2K$ and $2{\rm i}K'$ translations of their argument. The general solution can be put in the form:
\beq
\label{generalsolution}\omega(\psi) = \frac{1}{\nu(2 + \nn)} + g(\psi)\omega_+(\psi) + g(-\psi)\omega_+(-\psi - 2K),
\eeq
where $g(\psi)$ is an arbitrary analytic function which is $2K$ and $2{\rm i}K'$ periodic, and $\omega_+(\psi)$ is any analytic function satisfying:
\beq
\label{omegaplus}\omega_+(\psi + 2{\rm i}K') = \omega_+(\psi),\quad \omega_+(\psi + 2K) = e^{\pi\gamma}\omega_+(\psi),
\eeq
Here, the parametrization \eqref{ncosh} $\nn = -(e^{\pi\gamma} + e^{\pi\gamma})$ arises quite naturally. In \eqref{generalsolution}, the constant term is an obvious particular solution of the inhomogeneous equation \eqref{37}, while the second term is the general solution of \eqref{35a}-\eqref{37} with vanishing right-hand side.

Notice that, since $W$ was holomorphic on $\mathbb{C}\cup\{\infty\}\setminus[B_{\min},B_{\max}]$, and the right-hand side of \eqref{37} is a constant, $\omega(\psi)$ must be an entire function. Therefore, in the solution we are looking for, $g(\psi)\omega_+(\psi)$ (which also satisfies \eqref{omegaplus}) is entire as well. We explain in Appendix~\ref{appC} that this is impossible, unless:
\beq
\label{quanti}\gamma K' = 2pK\qquad\mathrm{for}\,\,\mathrm{some}\,\,\mathrm{positive}\,\,\mathrm{integer}\,\,p.
\eeq
In particular, $\gamma$ must be real, hence $c > 0$. Besides, when \eqref{quanti} holds, we easily guess a holomorphic solution of \eqref{omegaplus}:
\beq
\omega_+(\psi) = e^{\pi p \psi/K'}.
\eeq
With this choice, we are looking for an entire function $g(\psi)$ which is biperiodic, hence bounded, hence constant by Liouville theorem. We thus obtain\beq
\label{omega}\omega(\psi) = \frac{1}{\nu(2 + \nn)} + G\,\cosh[\pi(\gamma/2 + \psi/K')],
\eeq
for some constant $G$. The normalization \eqref{infinity} implies two consistency equations given by:
\beq
\label{normat}\omega(\psi) \mathop{=}_{\psi \rightarrow {\rm i}K'} 0 + \frac{N}{B_{\max}}(\psi - {\rm i}K') + o(\psi - {\rm i}K').
\eeq
Jointly with \eqref{quanti}, they provide three equations for the three unknowns $B_{\min}$, $B_{\max}$ and $G$. The integer $p$ seems to allow for a discrete set of solutions, but we now argue that $p = 1$ is the only one which is physically admissible. Indeed, let us come back to the density of poles \eqref{32}. After our change of variables, it can be expressed as:
\beq
\label{dens}\varrho(B) = \frac{\omega(K - {\rm i}\chi(B)K') - \omega(K + {\rm i}\chi(B)K')}{2{\rm i}\pi}
\eeq
where $0 < \chi(B) < 1$ is the unique point for which $\psi(B) = K + {\rm i}\chi(B)K'$ satisfies \eqref{paramtrigo} for $z = B$. Inserting \eqref{omega} into \eqref{dens}, we find:
\beq
\varrho(B) = -\frac{4G}{\pi}\,\frac{\sqrt{c}}{1 - c}\,\sin[\pi p \chi(B)]
\eeq
Since the sign of $\varrho(B)$ is not allowed to change on the support, we must impose $p = 1$.

The final answer for the resolvent can be written:
\bea
\label{ww}W(z(\psi)) & = & -\frac{1}{4\nu}\,\frac{(1 - c)^{3/2}}{c} \\
& & \times\Big\{\frac{1}{\sqrt{1 - c}} + \cosh[\pi(\gamma/2 + \psi/K')]\Big\}, \nonumber
\eea
and the corresponding pole density is:
\beq
\label{35}\varrho(B) = \frac{1}{2\pi}\,\sqrt{\frac{1 - c}{c}}\,\frac{1}{\nu}\,\sin[\pi\chi(B)].
\eeq
We can give an alternative definition for the function $\chi(B)$ in terms of Jacobi elliptic functions \cite{GuoWang}: if we set $k' = \sqrt{1 - k^2}$,
\beq
\label{36}\chi(B) = \frac{1}{K'}\,\nd^{-1}_{k'}\Big(\frac{B}{B_{\min}}\Big).
\eeq
The ratio $k = B_{\min}/B_{\max}$ is fixed as a function of $c = \tanh^2(\pi\gamma/2)$ by \eqref{quanti} with $p = 1$, and then we find from \eqref{normat}
\beq
\label{bmax}B_{\max} = \frac{8K}{\pi\gamma}\,\frac{\sqrt{c}}{1 - c}\,\nu N.
\eeq
To conclude, formula \eqref{slope} delivers the expression of the front slope:
\beq
\label{sll}\phi_x(x) = -\frac{1}{\sqrt{c}}\,\sin\Big[\frac{\pi}{K'}\,\mathrm{sc}^{-1}_{k'}\Big(\frac{x}{B_{\min}}\Big)\Big],
\eeq
where $\mathrm{sc}^{-1}_{k'}$ is the reciprocal function of the Jacobi elliptic function $\mathrm{sc}_{k'} = \mathrm{sn}_{k'}/\mathrm{cn}_{k'}$. These results are in agreement with the numerical resolution of \eqref{23}, see Fig.~\ref{Fi1}.

\medskip

\subsection{Solution for $-1 < c < 0$}

\label{aisa}
\begin{figure}
\includegraphics[width=8.7cm]{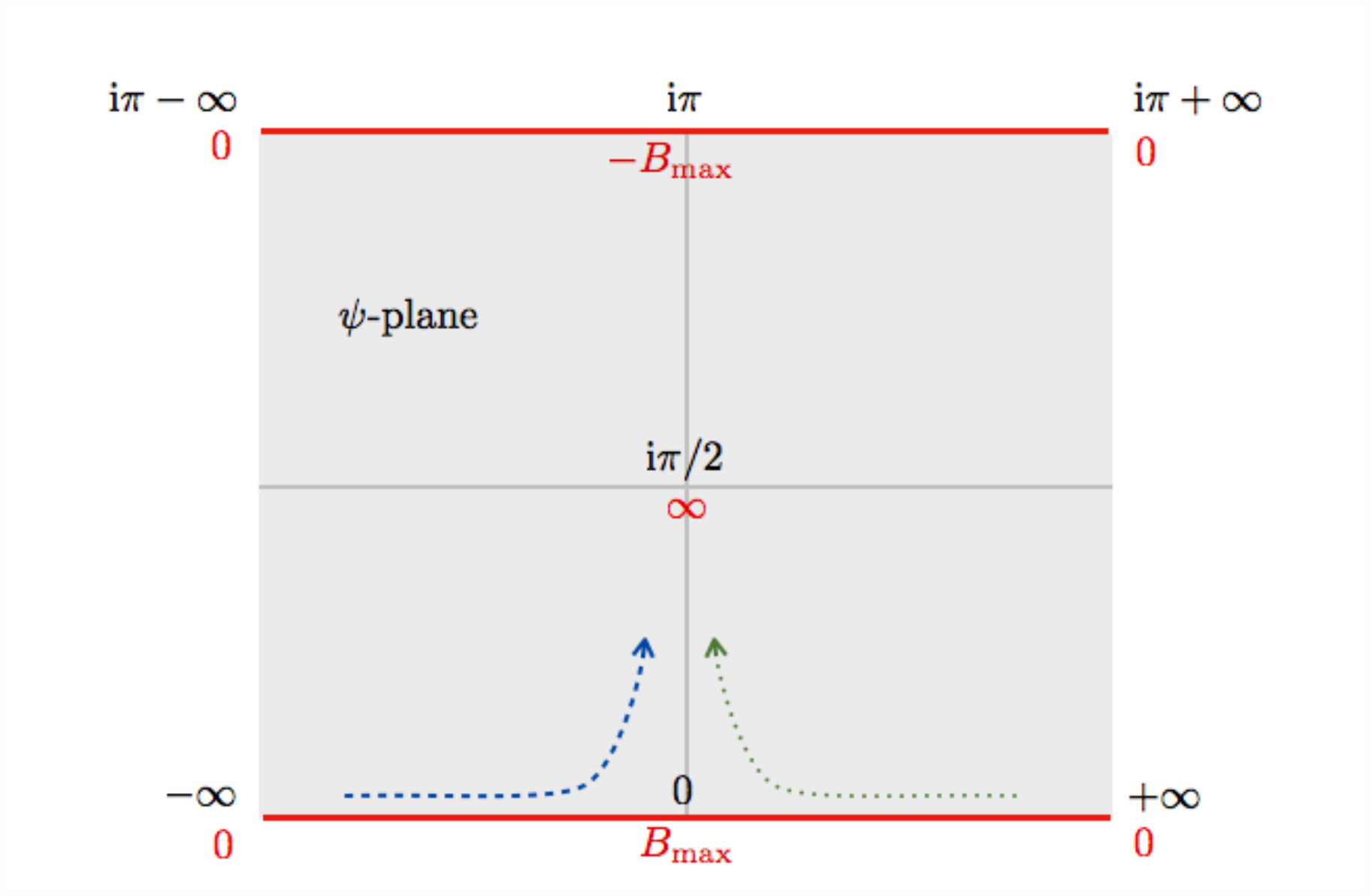}
\caption{\label{mappp} Mapping between the $z$-plane with two merged cuts $[-B_{\max},0]\cup[0,B_{\max}]$ and the strip $\mathcal{D}$ in the $\psi$-plane (values in black). For special points in the $\psi$-plane, we indicate below (in red) the corresponding value of $z$. The blue dashed (resp. green dotted) arrow indicates the image in the $\psi$-plane of a path in the $z$-plane with positive (resp. negative) imaginary part.}
\end{figure}

As we explained, the only possibility in this case is $B_{\min} = 0$. It corresponds to the limit $k \rightarrow 0$, and thus $K \rightarrow \pi/2$ in the former construction, in which the Jacobi elliptic function degenerates to trigonometric functions. We prefer to use the change of variable:
\beq
\label{tparam}z = \frac{B_{\max}}{\cosh \psi}.
\eeq
which, although not exactly the limit of \eqref{paramtrigo}, is closely related. Thanks to \eqref{tparam}, the $z$-complex plane with its two cuts $[-B_{\max},0]\cup[0,B_{\max}]$ is mapped onto the strip $\mathcal{D} = \{0 < \mathrm{Im}\,\psi < \pi\}$ (see Fig.~\ref{mappp}). More precisely, the segment $[0,B_{\max}] \pm {\rm i}0$ is mapped to the half-line $\mathbb{R}_{\mp}$ ; the segment $[-B_{\max},0] \pm {\rm i}0$ is mapped to the half-line ${\rm i}\pi + \mathbb{R}_{\mp}$ ; the point $z = \infty$ is mapped to $\psi = {\rm i}\pi/2$.

$\omega(\psi) = W(z(\psi))$ defines a holomorphic function on the strip $\mathcal{D}$. Since the analysis now parallels the case $c > 0$, we go quickly over it. Continuity of the resolvent $W(z)$ across $[-B_{\max},0]$ implies
\beq
\label{per}\omega(\psi) = \omega(2{\rm i\pi} - \psi)
\eeq
when ${\rm Im}\,\psi = \pi$, and \eqref{34} translates into
\beq
\label{per2}\omega(\psi) + \omega(-\psi) + \nn\,\omega(\psi + {\rm i}\pi) = 1/\nu
\eeq
when $\psi$ belongs to $\mathbb{R}$. These relations can be used to analytically continue $\omega(\psi)$ in the whole complex plane, so that \eqref{per}-\eqref{per2} are satisfied everywhere, or equivalently \eqref{per} and
\beq
\omega(\psi) + \omega(\psi + 2{\rm i}\pi) + \nn\,\omega(\psi + {\rm i}\pi) = 1/\nu.
\eeq
The general solution of \eqref{per}-\eqref{per2} can be written:
\beq
\label{omeg}\omega(\psi) = \frac{1}{\nu(2 + \nn)} + g(\psi)\,\omega_+(\psi) + g(-\psi)\,\omega_+(2{\rm i}\pi - \psi),
\eeq
where $g(\psi)$ is an arbitrary ${\rm i}\pi$ periodic, analytic function, and $\omega_+$ is any analytic function satisfying:
\beq
\omega_+(\psi + {\rm i}\pi) = e^{i\pi\gamma}\omega_+(\psi).
\eeq
where we recall the parametrization \eqref{ncos} $\nn = -(e^{{\rm i}\pi\gamma} + e^{-{\rm i}\pi\gamma})$. A possible choice is:
\beq
\label{omegap}\omega_+(\psi) = \frac{e^{(\gamma - 1)\psi}}{\sinh(\psi)}.
\eeq

Our change of variable \eqref{tparam} sends the point $z = 0$ at $\psi = \pm \infty$, and more precisely $z \sim (B_{\max}/2)\,e^{\mp\psi}$. So, we have to be careful that $W(\psi)$ does not grow too much when $\psi \rightarrow \pm \infty$ so that the pole density $\varrho(B)$ remains integrable at $B = 0$. This condition is equivalent  to demanding that $W(\psi)$ be $o(e^{\psi})$ when $\psi \rightarrow +\infty$. Being a ${\rm i}\pi$ periodic function, $g(\psi)$ admits a Fourier expansion $g(\psi) = \sum_{p \in \mathbb{Z}} \hat{g}_p\,e^{2p\psi}$. The integrability condition results in $\hat{g}_p = 0$ whenever $p \neq 0,1$. Besides, since the right-hand side of \eqref{per2} is a constant, we find that $\omega(\psi)$ is an entire function. We deduce that, in the solution we are looking for, $g(\psi)$ must be holomorphic, and such that the apparent pole of \eqref{omegap} at $\psi = {\rm i}\pi$ is absent in \eqref{omeg}. This gives a relation between $\hat{g}_0$ and $\hat{g}_1$, and leads eventually to our solution in the form:
\beq
\label{47}\omega(\psi) = \frac{1}{\nu(2 + \nn)} + G\,\cosh[\gamma(\psi - {\rm i}\pi)]
\eeq
for some constant $G$. The normalization condition \eqref{infinity} becomes:
\beq
\omega(\psi) \mathop{=}_{\psi \rightarrow {\rm i}\pi/2} 0 + \frac{{\rm i}N}{B_{\max}}(\psi - {\rm i}\pi/2) + o(\psi - {\rm i}\pi/2),
\eeq
and provides two equations determining the two unknowns $B_{\max}$ and $G$. We find that the maximal position of the poles is:
\beq
\label{bmax2}B_{\max} = \frac{4}{\gamma}\,\frac{\sqrt{-c}}{1 - c}\,N\nu.
\eeq
We thus find a unique solution for the resolvent:
\beq
\label{w}W(z(\psi)) = \frac{1}{4\nu}\,\frac{(1 - c)^{3/2}}{-c}\Big(\frac{1}{\sqrt{1 - c}} - \cosh[\gamma(\psi - {\rm i}\pi)]\Big),
\eeq
and as expected, the corresponding pole density is positive:
\beq
\label{varr}\varrho(B) = \frac{1}{2\pi}\,\sqrt{\frac{1 - c}{-c}}\,\frac{1}{\nu}\,\sinh\Big[\gamma\cosh^{-1}\Big(\frac{B_{\max}}{B}\Big)\Big],
\eeq
The flame slope is retrieved from \eqref{slope}:
\beq
\label{slop2}\phi_x(x) = - \frac{1}{\sqrt{-c}}\,\sinh\Big[\gamma\sinh^{-1}\Big(\frac{B_{\max}}{x}\Big)\Big].
\eeq
We observe a strong similarity with the results of the regime $0 < c < 1$, namely \eqref{ww}-\eqref{sll}. Going from $c > 0$ to $c < 0$, one just need to replace $c$ by $-c$, the elliptic functions by their trigonometric analog, and $\gamma$ is replaced by ${\rm i}\gamma$, so that trigonometric functions rather appears as hyperbolic functions.

\subsection{Discussion}

The first noteworthy point concerns $-1 < c < 0$: the results \eqref{bmax2} and \eqref{varr}-\eqref{slop2} then coincide with those found in \cite{joulindenet12}, up to slightly different notations (e.g., $\gamma$ was called $\mu$ whereby $\frac{2\sqrt{-c}}{\gamma(1 - c)} = \frac{\sin\pi\mu}{\mu}$) and thanks to such identities as $2\sinh^{-1}\big(\frac{1}{\sinh \xi}\big) = \mathrm{sgn}(\xi)\ln\big(\frac{1}{\tanh^2(\xi/2)}\big)$.

If $0 < c < 1$,	however,	the above exact results do differ from those found in \cite{joulindenet12} by naive continuation from $c < 0$. In particular, \eqref{sll} and \eqref{35}-\eqref{36} are free of the oscillations that crippled \eqref{12} and the corresponding density profile. To explain why the latter "solutions" were nevertheless so close to their numerical counterpart over the $x + {\rm i}B = O(N\nu)$ range when $0 \leq  c \lesssim 0.6$, one first notices that the maximum pole altitude $B_{\max}$ in \eqref{bmax} and the quantity $B_* = \frac{4}{\gamma}\,\frac{\sqrt{c}}{(1 - c)}\,\nu N$ featured in \eqref{12} are related by
\beq
\frac{B_{\max}}{B_*} = \frac{2K}{\pi} \mathop{=}_{k \rightarrow 0} 1 + O(k^2)
\eeq
As $K' = \ln(4/k) + o(1)$ when $k \rightarrow 0$, $\gamma \sim \frac{2\sqrt{c}}{\pi}$ must be small in \eqref{quanti} for $p = 1$, which implies:
\beq
\label{kexp}k \sim 4\exp\Big(-\frac{\pi^2}{2\sqrt{c}}\Big)
\eeq
The ratio $k = \frac{B_{\min}}{B_{\max}}$ is thus exceedingly small if $0 < c \ll 1$, and it remains small if $0 \leq c \lesssim 0.6$ thanks to the $\pi^2$ constant in \eqref{kexp}: $B_*$ and $B_{\max}$ are then nearly equal. Similarly, the difference between \eqref{sll} and \eqref{12} is strongly suppressed by $O(k^2)$ factors wherever $|x| \gg B_{\min}$ ; this is enough for \eqref{12} to accurately capture the crest slope profile, the exact maximum $\max |\phi_x(x)| = c^{-1/2}$ (reached at $|x|/B_{\min} \sim k^{-1/2} \gg 1$) inclusive.

The discarded unphysical profiles associated with integers $p > 1$ in \eqref{quanti} would have had an extra $p$ in the exponent of \eqref{kexp}. And, as $p \rightarrow \infty$ at fixed $0 < c < 1$ in \eqref{quanti} leads to $k \rightarrow 0$ and $B_{\min} \rightarrow 0$, the front slope $\phi_x(x)$ given by \eqref{sll} would resume \eqref{12}. Indeed, the only "error" in \eqref{12} was the implicit assumption that $B_{\min}/B_{\max} = 0$, as when $c < 0$.

Next, for $k \rightarrow 1^{-}$, we have $K' \sim \pi/2$ and $K = \frac{1}{2}\ln\big(\frac{8}{(1 - k)}\big) + o(1)$, while $\pi\gamma = \ln\big(\frac{4}{1 - c}\big) + o(1)$ if $c \rightarrow 1^{-}$ ; equations \eqref{quanti} and \eqref{bmax} then reduce to:
\beq
\Big(1 - \frac{B_{\min}}{B_{\max}}\Big) \sim 4\sqrt{1 - c},\qquad B_{\max} \sim \frac{2\nu N}{1 - c}
\eeq 
quite in line with Appendix~\ref{appA}. Using the integral representation of ${\rm nd}_{k'}^{-1}$ \cite[p. 494]{WW} in \eqref{35}-\eqref{36} and \eqref{kexp}, one can show that \eqref{35} resumes the Mar\u{c}enko-Pastur law when $c \rightarrow 1^{-}$.

\section{Bicoalesced periodic fronts}
\label{sec4}

\subsection{Integral equations for pole densities}
\label{sec41}
We now consider steady periodic cells, for which $\phi(t,x) = -Vt + \phi(x)$ ; the drift velocity was calculated in \cite{joulin91} in terms of the neutral-to-actual wavelength $\nu \in ]0,1[$ and the total number $N$ of pairs of poles involved $V = \frac{2N\nu(1 - N\nu)}{1 - c}$. Upon rescaling of the space coordinate, we may assume that $\phi_x(x)$ is $2\pi$-periodic. We will address more specifically the issue of bicoalesced periodic fronts. These are obtained as periodic solutions of the ZT equation which admit a pole decomposition, with two vertical piles of poles separated by a half-wavelength:
\beq
\label{41}\phi_x(x) = \sum_{\substack{k = -N_1 \\ k \neq 0}}^{N_1} \frac{-\frac{\nu}{1 - c}}{\tan\big(\frac{x - {\rm i}B_k}{2}\big)} + \sum_{\substack{m = -N_2 \\ m \neq 0}}^{N_2} \frac{-\frac{\nu}{1 - c}}{\tan\big(\frac{x - \pi - {\rm i}b_m}{2}\big)}
\eeq
A large number of these bicoalesced solutions have been obtained in \cite{denetstationary} in the Michelson-Sivashinsky case $c=0$, all of which have been found to be stable with Neumann boundary conditions in the same paper. A solution of this type, obtained by direct numerical simulation, is presented in \cite{Travnikov-et.al-00}. Periodic configurations featuring a single pile of pole can be extracted from \eqref{41} by assuming $N_2 = 0$, or assuming $N_1 = N_2$ and $b_k = B_k$, and doubling the wavelength. The analysis of such solutions is parallel to that of isolated crests. Plugging \eqref{41} in the ZT equation fixes the prefactor equal to $\frac{-\nu}{1 - c}$, and yields coupled equations for the position of the poles. In the first pile, we have for any $k = 1,\ldots,N_1$,
\bea
\label{42}\sum_{\substack{j = 1 \\ j \neq k}}^{N_1} \frac{1}{\tanh\big(\frac{B_k - B_j}{2}\big)} -\frac{\nn}{2} \sum_{j = 1}^{N_1} \frac{1}{\tanh\big(\frac{B_k + B_j}{2}\big)} & & \\
+ \sum_{m = 1}^{N_2} \tanh\big(\frac{B_k - b_m}{2}\big) -\frac{\nn}{2}\sum_{m = 1}^{N_2} \tanh\big(\frac{B_k + b_m}{2}\big) & = & f, \nonumber
\eea
and symmetrically in the second pile, we have for any $m = 1,\ldots,N_2$:
\bea
\label{43}\sum_{\substack{l = 1 \\ l \neq m}}^{N_2} \frac{1}{\tanh\big(\frac{b_m - b_l}{2}\big)} -\frac{\nn}{2} \sum_{l = 1}^{N_2} \frac{1}{\tanh\big(\frac{b_m + b_l}{2}\big)} & & \\
+ \sum_{k = 1}^{N_1} \tanh\big(\frac{b_m - B_k}{2}\big) -\frac{\nn}{2}\sum_{k = 1}^{N_1} \tanh\big(\frac{b_m + B_k}{2}\big) & = & f, \nonumber
\eea
where the constant in the right-hand side reads:
\beq
\label{56} f = \frac{1}{\nu}\Big(1 + \frac{2c}{1 - c}\,N\nu\Big),\qquad N = N_1 + N_2.
\eeq

We shall study \eqref{42}-\eqref{43} for large crests, i.e. in the limit $N_1,N_2 \rightarrow \infty$. We expect that the poles $B_1,\ldots,B_{N_1}$ of the first pile (resp. $b_1,\ldots,b_{N_2}$ of the second pile) get condensed on a segment $J_1 = [B_{\min},B_{\max}]$ (resp. $J_2 = [b_{\min},b_{\max}]$) with a continuous nonnegative density $\varrho_1(B)$ (resp. $\varrho_2(b)$), as discussed in Section~\ref{sec2}. The densities are normalized as:
\beq
\label{norm}\int_{J_1} \varrho_1(B)\,\dd B = N_1,\qquad \int_{J_2} \varrho_2(b)\,\dd b = N_2.
\eeq
and the flame slope \eqref{41} is retrieved as:
\beq
\label{fff}\phi_x(x) =  \frac{-2\nu}{1 - c}\,\mathrm{Re}\Big[\int_{J_1} \frac{\varrho_1(B')\,\dd B'}{\tan\big(\frac{x - {\rm i}B'}{2}\big)} + \int_{J_2}\frac{\varrho_2(b')\,\dd b'}{\tan\big(\frac{x - \pi - {\rm i}b'}{2}\big)}\Big]
 \eeq
The relations \eqref{42}-\eqref{43} then turn into coupled integral equations for the densities: for all $B \in J_1$, 
\bea
\label{46}\fint_{J_1} \frac{\varrho_1(B')\,\dd B'}{\tanh\big(\frac{B - B'}{2}\big)} - \frac{\nn}{2}\int_{J_1} \frac{\varrho_1(B')\,\dd B'}{\tanh\big(\frac{B + B'}{2}\big)} & & \\
+ \int_{J_2} \frac{\varrho_2(b')\,\dd b'}{\cotanh\big(\frac{B - b'}{2}\big)} - \frac{\nn}{2}\int_{J_2} \frac{\varrho_2(b')\,\dd b'}{\cotanh\big(\frac{B + b'}{2}\big)} & = & f,\nonumber 
\eea
and for all $b \in J_2$:
\bea
\label{47a}\fint_{J_2} \frac{\varrho_2(b')\,\dd b'}{\tanh\big(\frac{b - b'}{2}\big)} - \frac{\nn}{2}\int_{J_2} \frac{\varrho_2(b')\,\dd b'}{\tanh\big(\frac{b + b'}{2}\big)} & & \\
+ \int_{J_1} \frac{\varrho_1(B')\,\dd B'}{\cotanh\big(\frac{b - B'}{2}\big)} - \frac{\nn}{2}\int_{J_1}\frac{\varrho_1(B')\,\dd B'}{\cotanh\big(\frac{b + B'}{2}\big)} & = & f.\nonumber
\eea

Even though $N \rightarrow \infty$, the full expression of $f$ given in \eqref{56} is to be retained in \eqref{46}-\eqref{47a}. The reason why steady periodic fronts need an at least $O(N)$ actual-to-neutral wavelength ratio $1/\nu$ will be discussed in \S~\ref{dsis}.

\subsection{Reformulation by resolvents}

To take advantage of the periodicity, we perform a first change of variable:
\beq
\label{40}T = \frac{\tanh(B/2)}{\tanh(B_{\max}/2)},\qquad \tau = \frac{\tanh(b/2)}{\tanh(b_{\max}/2)}.
\eeq
and define the resolvents for a complex variable $z$:
\beq
\label{w1}W_1(z) = \int_{\tilde{J}_1} \frac{\varrho_1(B(T))\,\dd T}{z - T},\,\, W_2(z) = \int_{\tilde{J}_2} \frac{\varrho_2(b(\tau))\,\dd\tau}{z - \tau}.
\eeq
where $\tilde{J}_1 = [T_{\min},1]$ and $\tilde{J}_2 = [\tau_{\min},1]$ are the image of the supports after \eqref{40}. The normalization conditions \eqref{norm} are now rephrased as:
\bea
N_1 & = & \int_{\tilde{J}_1} \frac{2\tanh(B_{\max}/2)\varrho_1(B(T))\,\dd T}{1 - \tanh^2(B_{\max}/2)T^2} \nonumber \\
& = & \oint_{\tilde{J}_1} \frac{\dd z}{2{\rm i}\pi} \frac{2\tanh(B_{\max}/2)\,W_1(z)}{1 - \tanh^2(B_{\max}/2)z^2},
\eea
and the contour of integration can be moved at infinity to pick up residues at $z = \pm 1$. Therefore:
\beq
\label{norma}N_1 = W_1\Big(\frac{1}{\tanh(B_{\max}/2)}\Big) - W_1\Big(\frac{-1}{\tanh(B_{\max}/2)}\Big).
\eeq
The same computation holds for $N_2$, with $W_1$ and $B_{\max}$ replaced in \eqref{norma} by $W_2$ and $b_{\max}$. Starting from the essential properties of resolvents emphasized in \eqref{32}-\eqref{333}, the relations \eqref{46}-\eqref{47a} can be transformed after a tedious algebra into a Riemann-Hilbert problem for $W_i$: for any $T \in \tilde{J}_1$,
\bea
W_1(T + {\rm i}0) + W_1(T - {\rm i}0) + \nn\,W_1(-T) & & \nonumber \\
\label{44}+ 2W_2(1/PT) + \nn\,W_2(-1/PT) & = & \tilde{f},
\eea
and a symmetric equation for any $\tau \in \tilde{J}_2$,
\bea
W_2(\tau + {\rm i0}) + W_2(\tau - {\rm i}0) + \nn\,W_2(-\tau) & & \nonumber \\
\label{45}+ 2W_1(1/P\tau) + \nn\,W_2(-1/P\tau) & = & \tilde{f},
\eea
where we have introduced the -- yet unknown -- constants:
\bea
\label{coc}P & = & \tanh(B_{\max}/2)\tanh(b_{\max}/2), \\
\label{68} \tilde{f} & = & f - \frac{4c}{1 - c}\,M,  \\
M & = & \int_{\tilde{J}_1} \frac{\tanh^2(B_{\max}/2)\,T\,\varrho_1(B(T))\dd T}{1 - \tanh^2(B_{\max}/2)\,T^2} \nonumber \\
& & + \int_{\tilde{J}_2} \frac{\tanh^2(b_{\max}/2)\,\tau\,\varrho_2(b(\tau))\dd \tau}{1 - \tanh^2(b_{\max}/2)\,\tau^2}.
\eea 

\subsection{Symmetric case}
\label{sum}
As of writing, we could not solve \eqref{44}-\eqref{45} in their full generality. Yet, we did determine the solution when there is some symmetry between the two piles of poles, the same as for $c = 0$ \cite{joulindenet08}, \textit{viz.}:
\beq
\label{symr}\tilde{J}_1 = \tilde{J}_2 \equiv [r,1]\,\,\mathrm{and}\,\,\varrho_1(B(T)) = \varrho_2(b(T)) \equiv \varrho(T).
\eeq
This means that $\varrho_1(B)$ and $\varrho_2(b)$ can be deduced from one another by the rescaling encoded in \eqref{40}. We stress that in general $\varrho_1(B) \neq \varrho_2(B)$, unless $B_{\max} = b_{\max}$, i.e. unless they have the same support (in which case \eqref{fff} describes two copies of a monocoalesced periodic cell of length $\pi$). Likewise, the number of poles in each pile need not be equal. When \eqref{symr} is satisfied, we obviously have $W_1(z) = W_2(z)$. The trick is now to introduce a new resolvent, which will incorporate simultaneously the terms involving $z$ and $1/z$ in \eqref{44}-\eqref{45}. For this purpose, we perform a second change of variable:
\beq
\label{eta}\eta = \frac{T}{1 + PT^2},
\eeq
already introduced in \cite{joulindenet12} to deal with the case $c = 0$, and which is invariant under $T \leftrightarrow 1/PT$. Though not invertible on the whole complex $T$-plane, \eqref{eta} sends bijectively the support $[r,1]$ of the density $\varrho$ in the $T$-plane, to the segment:
\beq
\label{sai}\hat{J} = \big[\frac{r}{1 + P r^2},\frac{1}{1 + P}\big] \equiv [\eta_{\min},\eta_{\max}]
\eeq
in the $\eta$-plane. We can thus define:
\beq
\label{red}\hat{\rho}(\eta) = \varrho(T(\eta)),
\eeq
which is a density supported on $\hat{J}$, and next the resolvent for a complex variable $z$:
\beq
\label{hatW}\hat{W}(z) = \int_{\hat{J}} \frac{\hat{\rho}(\eta)\,\dd\eta}{z - \eta}.
\eeq
A computation shows that it is related to \eqref{w1} by:
\beq
\label{haha}\hat{W}\big(\frac{z}{1 + Pz^2}\big) = W(z) + W(1/Pz) + \int_{r}^{1} \frac{2PT\,\varrho(T)\dd T}{1 + PT^2}.
\eeq
We deduce from \eqref{44} (or \eqref{45}) the relation, for all $\eta \in \hat{J}$:
\beq
\label{loop}\hat{W}(\eta + {\rm i}0) + \hat{W}(\eta - {\rm i}0) + \nn\,\hat{W}(-\eta) = \hat{f}.
\eeq
The constant in the right hand-side combines \eqref{coc} and the last term in \eqref{haha}, and becomes quite simple at the end:
\beq
\hat{f} = f - \frac{4c}{1 - c}\,\hat{M},
\eeq
where
\bea
\hat{M} & = &  \int_{\hat{J}} \frac{S^2\eta\,\hat{\rho}(\eta)\dd\eta}{1 - S^2\eta^2} = \frac{1}{2}\big(\hat{W}\big(1/S) + \hat{W}(-1/S)\big), \nonumber \\
\label{coco} S & = & \tanh(B_{\max}/2) + \tanh(b_{\max}/2),
\eea
and we have evaluated $\hat{M}$ thanks to Cauchy residue formula. We now recognize an equation of the same type as \eqref{34}, hence solvable. The main difference with Section~\ref{sec2} is that the data of the number of poles $N_1$ and $N_2$ are encoded in a different way in the resolvent. We can rewrite \eqref{norma} in terms of $\hat{\rho}$ only:
\beq
N_i = \int_{\hat{J}} \Big(S + \frac{\epsilon_i\,D}{\sqrt{1 - 4P\eta^2}}\Big)\frac{\hat{\rho}(\eta)\dd\eta}{1 - S^2\eta^2},\qquad i = 1,2,
\eeq
where $\epsilon_1 = +1$, $\epsilon_2 = -1$, and:
\beq
D = \tanh(B_{\max}/2) - \tanh(b_{\max}/2)
\eeq
So, the total number of poles is retrieved in a simple way from $\hat{W}$:
\beq
\label{79}N = \hat{W}(1/S) - \hat{W}(-1/S),
\eeq
whereas the difference between the two piles is only given implicitly:
\beq
\label{80}\frac{N_1 - N_2}{2} = \int_{\hat{J}} \frac{D}{\sqrt{1 - 4P\eta^2}}\,\frac{\hat{\rho}(\eta)\dd\eta}{(1 - S^2\eta^2)}.
\eeq
Eventually, the flame slope \eqref{fff} can be expressed directly in terms of the resolvent: if we first define
\beq
\label{sdef}s(x) = \frac{\tan(x/2)}{\tanh(B_{\max}/2) - \tanh(b_{\max}/2)\tan^2(x/2)},
\eeq
we find after some algebra that:
\bea
\phi_x(x) & = &  -\frac{4\nu}{1 - c} \int_{\hat{J}} \frac{s(x)\,\hat{\rho}(\eta)\dd\eta}{\eta^2 + s^2(x)} \nonumber \\
\label{83} & = & \frac{4\nu}{1 - c}\,\mathrm{Im}\big[\hat{W}({\rm i}s(x))\big].
\eea
Before coming to the determination of $\phi_x(x)$, let us recapitulate the unknowns and the parameters for those symmetric solutions.
Initially, the supports $[B_{\min},B_{\max}]$ and $[b_{\min},b_{\max}]$ of the pole densities in the first and second pile were unknown. The symmetric case we are studying amounts to consider supports with the same aspect ratio, namely:
\beq
r = \frac{\tanh(B_{\min}/2)}{\tanh(B_{\max}/2)} = \frac{\tanh(b_{\min}/2)}{\tanh(b_{\max}/2)}.
\eeq
We are left with two unknowns, which can be encoded in the variables $P$ and $S$ defined in \eqref{coc} and \eqref{coco}. On the other hand, $c$ (or equivalently $\nn$ or $\gamma$), and the number of poles in each pile $N_1$ and $N_2$, count as free parameters.

\subsection{Solution for $0 < c < 1$}
\label{soa}
\begin{figure}
\includegraphics[width=8.5cm]{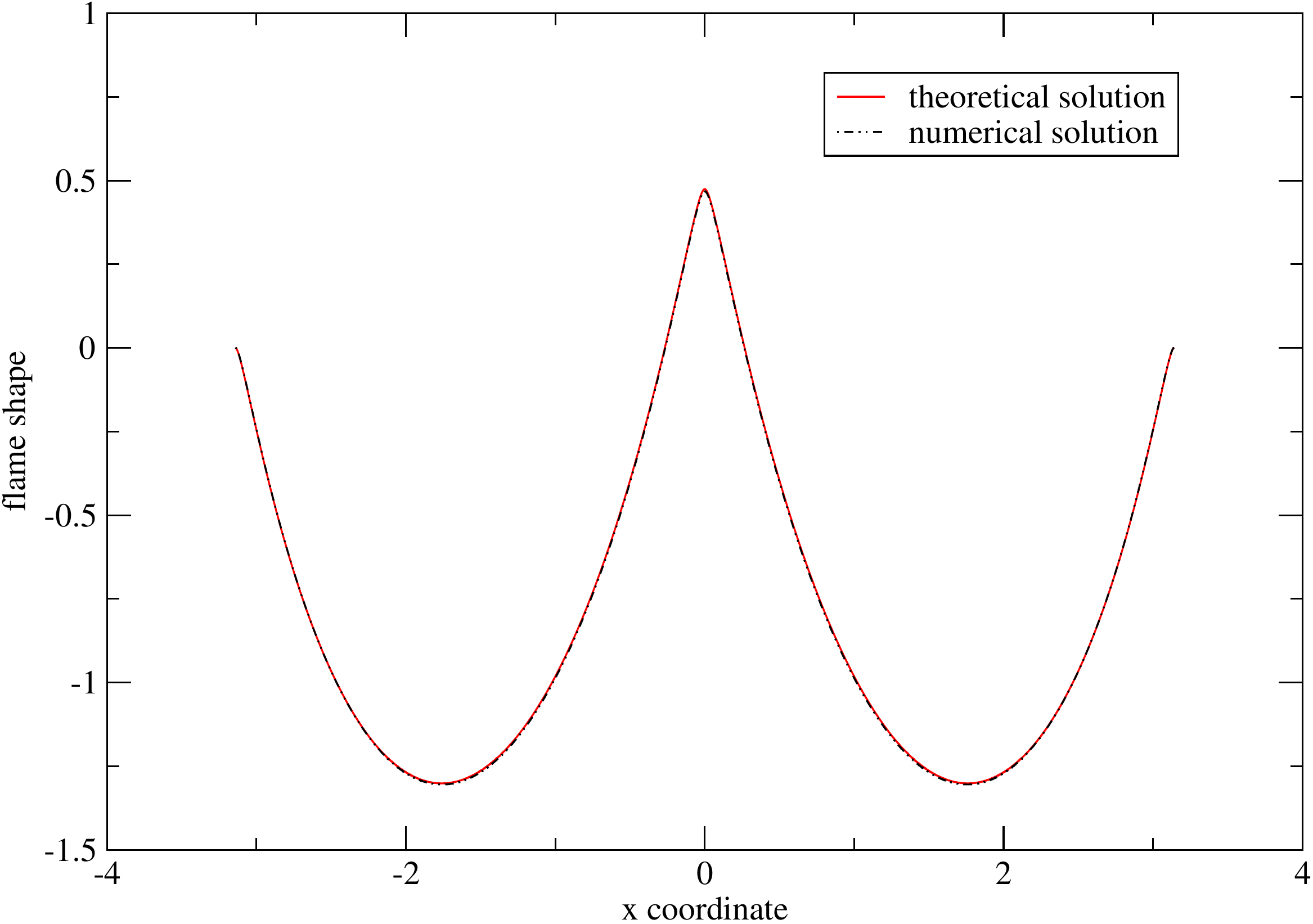}
\caption{\label{F2} Bicoalesced periodic front shape $\phi(x)$ with $c > 0$. The plot assumes $(N_1,N_2) = (200,100)$, $1/\nu = 600.5$ and $c = 0.5$, We display the comparison between the numerical resolution of \eqref{42}-\eqref{43} and integration of the theoretical formula \eqref{99}.}
\end{figure}

If we assume $B_{\min},b_{\min} > 0$, we have $\eta_{\min} > 0$ (see \eqref{sai}). We can repeat the analysis of Section~\ref{sec31}. We first perform the change of variable
\beq
\label{etan}z = \eta_{\min}\,\sn_{k}\psi,\qquad k = \frac{\eta_{\min}}{\eta_{\max}} = \frac{(1 + P)r}{1 + Pr^2},
\eeq
and $\omega(\psi) = \hat{W}(z(\psi))$ can be extended as a holomorphic function of $\psi$ in the whole complex plane. We deduce that $\eta_{\min} > 0$ implies $c > 0$, and the quantization condition:
\beq
\label{quantiq}\gamma K' = 2pK\qquad\mathrm{for}\,\,\mathrm{some}\,\,\mathrm{positive}\,\,\mathrm{integer}\,\,p.
\eeq
We already know that the positivity of the pole density requires $p = 1$. Besides, the general solution of \eqref{loop} in terms of $\omega$ is obtained from \eqref{omega}, where $1/\nu$ (which appeared in the right-hand side of \eqref{34}) is replaced by $\hat{f}$ given in \eqref{56} and \eqref{68}. Exploiting the parametrization \eqref{nc} of $\nn$ in terms of $c$, we arrive at:
\bea
\label{omega2}
\omega(\psi) & = & \frac{f}{2 + \nn} + \frac{1}{2}\big[\omega(\psi_{1/S}) + \omega(-\psi_{1/S})\big] \nonumber \\
& & + G\,\cosh\big[\pi(\gamma/2 + \psi/K')\big],
\eea
where $G$ is a constant to determine and $\psi_{1/S} = {\rm i}K' + \delta$ with $0 < \delta < K$ is the unique point satisfying \eqref{etan} for $z = 1/S$. This can be rephrased as:
\beq
\label{92}\sn_k(\delta) = \frac{S}{1 + P} = \tanh\Big(\frac{B_{\max} + b_{\max}}{2}\Big).
\eeq
We have a priori five unknowns: $S,P,r$ as discussed in Section~\ref{sum}, the constant $G$, and $\frac{1}{2}\big[\omega(\psi_{1/S}) + \omega(-\psi_{1/S})\big]$ (which is in fact irrelevant in the final result for the density and the flame slope) appearing in \eqref{omega2}. We already have two normalization conditions \eqref{79}-\eqref{80} and the quantization condition \eqref{quantiq}. A fourth equation comes from the fact that $\hat{W}(\eta = \infty) = 0$ (see \eqref{hatW}), hence $\omega({\rm i}K') = 0$, and fixes the value of $\frac{1}{2}\big[\omega(\psi_{1/S}) + \omega(-\psi_{1/S})\big]$ in terms of the other unknowns. The fifth and last equation is obtained by consistency of \eqref{omega2}: if we specialize to $\psi = \pm\psi_{1/S}$ and consider the half-sum of the two equations, we find:
\beq
\label{qqq}- G\cosh(\pi\gamma/2)\cosh\Big(\frac{\pi\delta}{K'}\Big) = -\frac{f}{2 + \nn}.
\eeq
Besides, \eqref{79} can be rewritten:
\beq
\label{qqa2}- 2G\sinh(\pi\gamma/2)\sinh\Big(\frac{\pi\delta}{K'}\Big) = N.
\eeq
Solving for $G$ and $\psi_{1/S}$ in the system \eqref{qqq}-\eqref{qqa2}, and taking into account the expression \eqref{56} for $f$, yields:
\beq
\label{Fdef}G = -\frac{F}{4\nu}\,\frac{(1 - c)^{3/2}}{c},\qquad F = \sqrt{1 + \frac{4c}{1 - c}N\nu(1 - N\nu)},
\eeq
and:
\beq
\label{del} \frac{S}{1 + P} = \sn_k\Big[\frac{K'}{\pi}\,\tanh^{-1}\Big(\frac{2N\nu\,\sqrt{c}}{1 - c(1 - 2N\nu)}\Big)\Big] .
\eeq
Then, the system of equations \eqref{80},\eqref{quantiq},\eqref{del} determines implicitly $S, P$ and $r$ in terms of $N_1,N_2$ and $c$, and it can be solved at least numerically. The final answer for the reduced density of poles \eqref{symr},\eqref{red} reads:
\beq
\label{97}\hat{\varrho}(\eta) = \frac{1}{2\pi}\,\sqrt{\frac{1 - c}{c}}\,\frac{F}{\nu}\,\sin[\pi\chi(\eta)],
\eeq
where we have introduced:
\beq
\chi(\eta) = \frac{1}{K'}\,\nd_{k'}^{-1}\Big(\frac{1 + Pr^2}{r}\,\eta\Big),\qquad k' = \sqrt{1 - k^2}.
\eeq
And from \eqref{83}, we deduce the flame slope:
\beq
\label{99}\phi_x(x) = -\frac{F}{\sqrt{c}}\,\sin\Big[\frac{\pi}{K'}\,\mathrm{sc}_{k'}^{-1}\Big(\frac{(1 + Pr^2)}{r}\,s(x)\Big)\Big],
\eeq
where $s(x)$ and $F$ were defined in \eqref{sdef} and \eqref{Fdef}. A flame shape $\phi(x)$ corresponding to \eqref{99} is plotted in Fig.~\ref{F2}.

\subsection{Solution for $-1 < c < 0$}
\label{sec44}
\begin{figure}
\includegraphics[width=8.5cm]{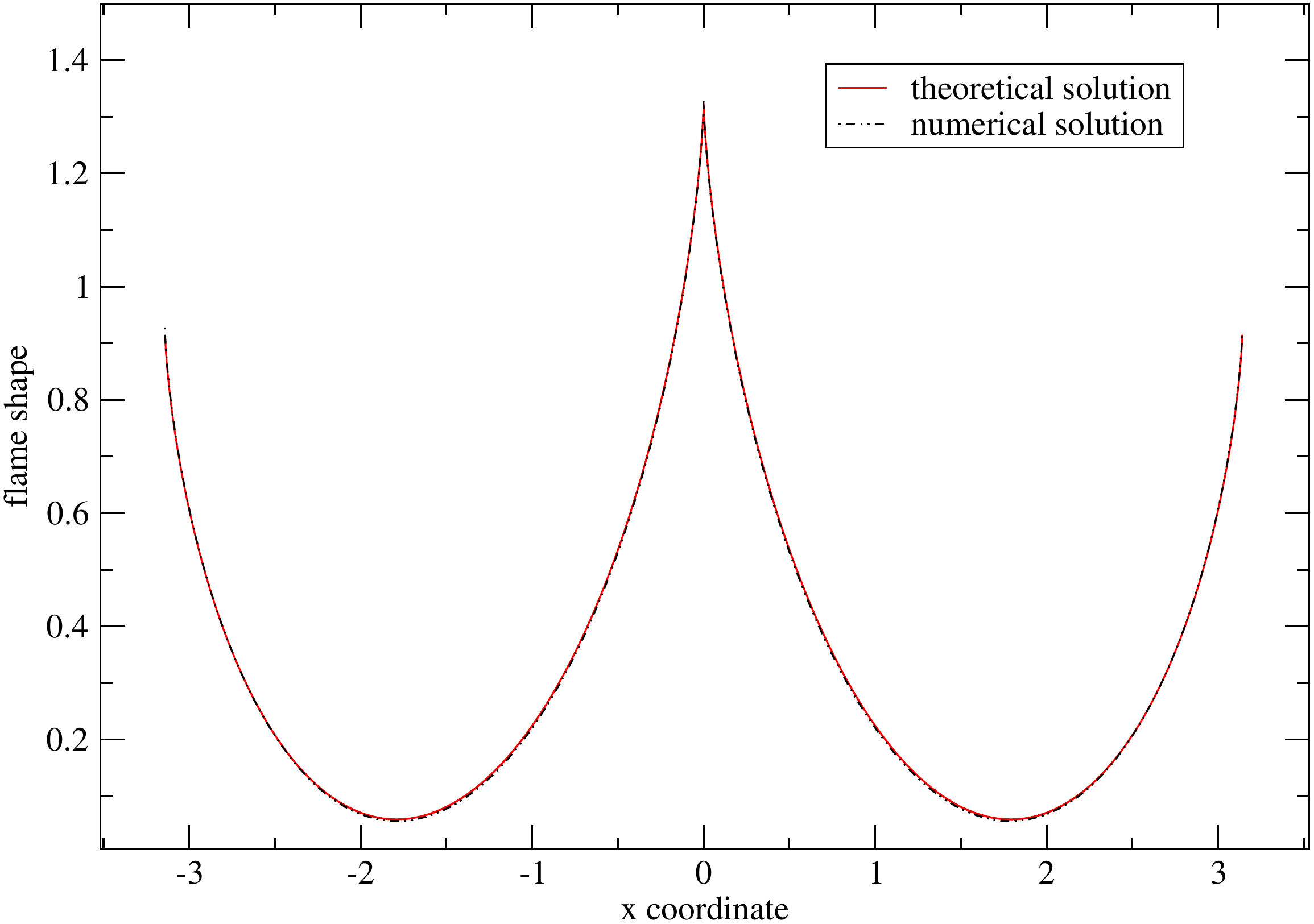}
\caption{\label{F3} Bicoalesced periodic front shape $\phi(x)$ with $c < 0$. The plot assumes $(N_1,N_2) = (200,100)$, $1/\nu = 600.5$ and $c = -0.25$. We display the comparison between the numerical resolution of \eqref{42}-\eqref{43} and integration of the theoretical formula \eqref{106}.}
\end{figure}

In this regime, we must have $B_{\min} = b_{\min} = 0$, i.e. $\eta_{\min} = 0$ or equivalently $r = 0$, and we can repeat the analysis of Section~\ref{aisa}. We first perform the change of variable:
\beq
\label{aq}z = \frac{\eta_{\max}}{\cosh \psi},
\eeq
the function $\omega(\psi) = \hat{W}(z(\psi))$ can be extended as a holomorphic function in the whole $\psi$-plane, and has the general expression \eqref{47} where $1/\nu$ is replaced -- as in Section~\ref{soa} -- by the appropriate constant term:
\bea
\omega(\psi) & = & \frac{f}{2 + \nn} + \frac{1}{2}\big[\omega(\psi_{1/S}) + \omega({\rm i}\pi -\psi_{1/S})\big] \nonumber \\
& & + G\cosh[\gamma(\psi - {\rm i}\pi)],
\eea
where $G$ is some constant to determine, and $\psi_{1/S} = {\rm i}(\pi/2 - \delta)$ with $0 < \delta < \pi/2$ is the unique point satisfying \eqref{aq} for $z = 1/S$. In other words:
\beq
\label{102}\sin(\delta) = \frac{S}{1 + P} = \tanh\Big(\frac{B_{\max} + b_{\max}}{2}\Big).
\eeq
Since the discussion to determine the unknowns is also very similar to Section~\ref{soa}, we only give the results:
\beq
G = -\frac{F}{4\nu}\,\frac{(1 - c)^{3/2}}{-c}
\eeq
with the same constant $F$ appearing in \eqref{Fdef}, and:
\beq
\label{888}\frac{S}{1 + P} = \sin\Big[\frac{1}{\gamma}\tan^{-1}\Big(\frac{2N\nu\,\sqrt{-c}}{1 - c(1 - 2N\nu)}\Big)\Big].
\eeq
Together with the implicit relation \eqref{80}, Eq.~\eqref{888} fixes $S$ and $P$, and thus the solution of our problem. The final result for the reduced density of poles \eqref{symr},\eqref{red} reads:
\beq
\label{105}\hat{\varrho}(\eta) = \frac{1}{2\pi}\sqrt{\frac{1 - c}{-c}}\,\frac{F}{\nu}\,\sinh\Big[\gamma\cosh^{-1}\Big(\frac{1}{(1 + P)\eta}\Big)\Big],
\eeq
and the corresponding flame slope is:
\beq
\label{106}\phi_x(x) = -\frac{F}{\sqrt{-c}}\,\sinh\Big[\gamma\sinh^{-1}\Big(\frac{1}{(1 + P)s(x)}\Big)\Big],
\eeq
where $s(x)$ and $F$ were defined in \eqref{coc} and \eqref{Fdef}. As expected, the resulting expression for $\phi_x(x)$ is symmetric under $(x,B_{\max}) \leftrightarrow (x - \pi,b_{\max})$. A flame shape $\phi(x)$ corresponding to \eqref{106} is plotted in Fig.~\ref{F3}.

\subsection{Discussion}
\label{dsis}

As is patent on comparing \eqref{105}-\eqref{106} with \eqref{97}-\eqref{99}, using $-1 < c < 0$ instead of $0 < c < 1$ again merely amounts to selecting the relevant expression of $\gamma$, and to replacing the elliptic functions by suitable hyperbolic ones, in the pole density and flame slope.  Both situations thus share several trends, which are appropriate to discuss first.

Since  $\sn_{k}(K) = 1 = \sin(\pi/2)$, $B_{\max} + b_{\max} \rightarrow \infty$ follows in the limit $2\nu N \rightarrow 1$, from \eqref{etan}, \eqref{92} and \eqref{del} if $0 < c < 1$, or \eqref{102} and \eqref{888} if $-1 < c < 0$. Next, \eqref{80} indicates that only the uppermost end of the most populated pile (namely, $B_{\max}$ if $N_1 > N_2$) actually goes to infinity in either case ; since $B_{\max}$ cannot go further vertically, steady bicoalesced periodic patterns cannot support an arbitrary large number of poles per wavelength \cite{joulin91}, just like when $c = 0$ \cite{TFH85,joulindenet08}. This can be traced back to \eqref{42}-\eqref{43}, for example with $b_{N_2} < B_{N_1} \rightarrow \infty$ if $N_1 > N_2$, in which limit the condition
\beq
N = N_1 + N_2 = N_{\mathrm{opt}}(\nu) = \lfloor \frac{\nu + 1}{2\nu} \rfloor
\eeq
is obtained. Note that $N_{\mathrm{opt}}(2\nu) \leq N_{\mathrm{opt}}(\nu)/2$, but the difference is asymptotically small for large wrinkles, as $2\nu N_{\mathrm{opt}}(\nu) \sim 1$  for $\nu \rightarrow 0^{+}$. Conversely, a large actual-to-neutral wavelength ratio $1/\nu \geq (2N - 1)$ is required for a total population of $2N$ bicoalesced poles to stay steady "in" a cell. 

\begin{figure}
\includegraphics[width=8.5cm]{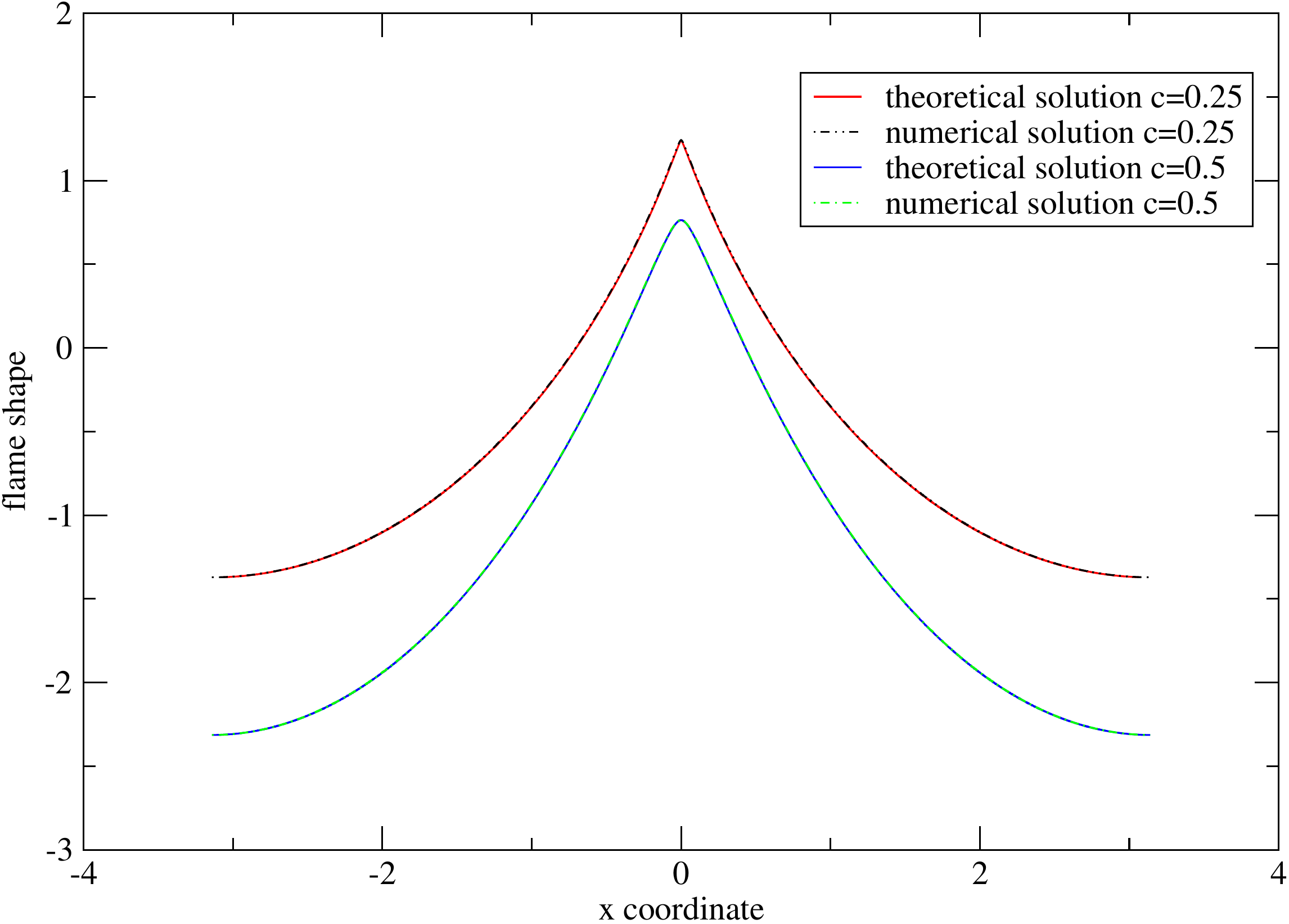}
\caption{\label{F1} Monocoalesced periodic front shapes $\phi(x)$ with $c > 0$. The plots assume $(N_1,N_2) = (100,0)$ and $1/\nu = 199.5$. For the two values $c = 0.25$ (top) and $c = 0.5$ (bottom), we show the comparison between the numerical resolution of \eqref{42}-\eqref{43} and integration of the theoretical formula \eqref{99}.}
\end{figure}

The analysis of Section~\ref{sec4} of the large steady wrinkles governed by the ZT equation \eqref{11} contain as limiting cases all the results available so far. Monocoalesced $2\pi$-periodic cells correspond to $2\nu N_2 \rightarrow 0$ at fixed $\nu N_1$ (see Fig~\ref{F1}). In the present continuous approximation of the pole equations, $b_{\max}$ then vanishes and this amounts to replacing:
\bea
\frac{1}{(1 + P)\eta} & \longrightarrow & \frac{\tanh(B_{\max}/2)}{\tanh(B/2)}, \nonumber \\
\label{107} \frac{1}{(1 + P)s(x)} & \longrightarrow & \frac{\tanh(B_{\max}/2)}{\tan(x/2)},
\eea
in \eqref{105} and \eqref{106} respectively. For such two-crested periodic patterns, $\frac{\tanh(B_{\max}/2)}{\tan(x/2)}$ plays the part $\frac{B_{\max}}{x}$ did for isolated crests, see \eqref{35}-\eqref{sll} and \eqref{varr}-\eqref{slop2}. Also, if $-1 < c < 0$, \eqref{888}-\eqref{106} then resume the expressions found in \cite{joulindenet08}. The isolated crests constitute a further degeneration of \eqref{107} and formally corresponds to $2\nu N_1 \ll 1$ and restricting oneself to a region of the complex plane where $|x + {\rm i}B| = O(N_1\nu) \ll 1$. Then $B_{\max} = O(N_1\nu) \ll 1$ by \eqref{92} or \eqref{102},
\beq
\frac{1}{(1 + P)\eta} \rightarrow \frac{B_{\max}}{B},\qquad\frac{\tanh(B_{\max}/2)}{\tan(x/2)} \rightarrow \frac{B_{\max}}{x},
\eeq
in \eqref{107}, the prefactor $F$ defined in \eqref{Fdef} goes to $1$, and the density and the flame slope reduce to \eqref{35}-\eqref{sll} or \eqref{varr}-\eqref{slop2}.

For equally populated piles of poles (i.e. $N_1 = N_2$), we have $b_{\max} = B_{\max}$, whence \eqref{eta} and \eqref{sdef} simplify to:
\beq
\label{109}\frac{1}{(1 + P)s(x)} = \frac{\tanh B_{\max}}{\tan x},\qquad \frac{1}{(1 + P)\eta} = \frac{\tanh B_{\max}}{\tanh B}.
\eeq
The comparison with \eqref{107} shows that \eqref{109} describes two copies of a monocoalesced cell of wavelength $\pi$: this was actually expected from the outset in view of the discrete pole equations \eqref{42}-\eqref{43} with $N_1 = N_2$, since $\tanh(a) + \frac{1}{\tanh(a)} = \frac{2}{\tanh(2a)}$.
 
The main qualitative difference  between the  $c \leq 0$ and $c > 0$ cases undoubtedly is about the tips of large front wrinkles. Even when $N_{1,2} \rightarrow \infty$ (and keeping $\nu N_{1,2} = O(1)$),  these are smooth maxima of $\phi(x)$ if $c > 0$, with $O(B_{\min})$ radii of curvature that quickly increase with $c$ if $0 < c \ll 1$ (see Fig.~\ref{F1}-\ref{F2}), and are sharp cusps otherwise (see Fig.~\ref{F3}). Specifically, the centered tips locally have $\phi_x(x) \propto -\mathrm{sgn}(x)\,|x|^{-\alpha}$ and thus
\beq
\phi(x) - \phi(0) \propto -|x|^{1 - \alpha},\qquad \alpha = \frac{2}{\pi}\tanh^{-1}(\sqrt{-c}).
\eeq
with $0 < \alpha < 1/2$ for $-1 < c < 0$. The MS value $c = 0$ is marginal and was already known to give logarithmic cusps \cite{TFH85,joulindenet08}. While affecting the crests ($-\mathcal{H}[\phi_x] > 0$), the value of $c$ also modifies the front troughs (minima of $\phi(x)$, where $\mathcal{H}[\phi_x] > 0$), but differently. This can be qualitatively understood from \eqref{11}, by transferring the nonlocal nonlinearity $c\mathcal{H}[\phi_x]^2$ to the right-hand side : if $c < 0$, the resulting "effective driving term" of instability, $-\mathcal{H}[\phi_x](1 + \frac{c}{2}\mathcal{H}[\phi_x])$, is reduced (or enhanced) at troughs (or crests) compared to its MS counterpart alone and tends to render them rounder (or sharper) ; $c > 0$ has opposite influences.

And, importantly again, whatever the value of $-1 \leq c < 1$ is, the analytical predictions of Section~\ref{sec4} perfectly agree with the numerical resolutions of \eqref{42}-\eqref{43} for the discrete poles $B_{k}$ and $b_m$, and with the bicoalesced front slopes ensuing from \eqref{41} and numerical quadrature. In particular, $\phi(x)$ is found for $c > 0$ to have 4 inflexion points whenever $N_1/N_2 \neq 0,\infty$ (see Fig.~\ref{F2}), and only 2 inflexion points otherwise ; in either case, $\max |\phi_x(x)| = \frac{F}{\sqrt{c}}$ is the same at all of them and only depends on $c > 0$ and $N\nu$.

\section{Conclusion}
\label{sec5}

We have investigated with complex analysis methods solutions of the ZT equation with free parameter $-1 < c < 1$, describing stationary flame shapes in the limit of large wrinkles. We rederived the known expression \eqref{slop2} for a front with an isolated crest when $-1 < c < 0$, and obtained its counterpart \eqref{sll} in the regime $0 < c < 1$. We also obtained new expressions describing bicoalesced, space-periodic fronts with some symmetry (see \S~\ref{sum}) in the regime $0 < c < 1$ \eqref{99} or $-1 < c < 0$ \eqref{106}, which are in agreement with numerical simulations (resp. Fig.~\ref{F2} and Fig.~\ref{F3}). The results obtained in the limit case of monocoalesced periodic fronts are also new (see Section~\ref{dsis}, and Fig~\ref{F1} for comparison to numerics). 

The preceding analyses do not exhaust all the theoretical problems as to \eqref{11}, for it admits even more general solutions than the symmetric bicoalesced ones studied in Section~\ref{sec4}. For example, the so-called \emph{interpolating solutions} \cite{GuidiMarchetti,denetstationary} are still awaiting for detailed descriptions. Their existence can be inferred on noticing that, besides $2N$ bicoalesced poles vertically aligned at $x = 0$ or $x = \pi$, $N_\infty$ extra pairs may stay in equilibrium at $Z_k = \pm {\rm i}\infty + \frac{(2k - 1)\pi}{N_\infty}$ for $k = 1,\ldots,N_{\infty}$ if $0 \leq 2\nu N = 1 - \nu N_{\infty} < 1$ ; with $2\nu N$ being further reduced, the $N_\infty$ remote poles will move and be located at finite $\mathrm{Im}\,Z_k$ \cite{GuidiMarchetti,denetstationary}. Although such solutions are likely unstable (as in the $c = 0$ case), their computation in the limit $\nu \rightarrow 0^+$ and still $\nu N_{1,2,\infty} = O(1)$ with similar methods is under investigation. As another example, space-periodic configurations with more than 2 unequal piles per cell certainly exist if $\nu$ is small: a single pair of poles may already stay in equilibrium near the trough of a base front if wide enough, thereby creating an incipient extra crest there \cite{joulindenet08}.  Whether this is within reach of resolvent approaches when the extra poles get many constitutes a challenge still to be met, as some of the poles would again condense on curved arcs in such configurations.
  
In the continuous description of, say, isolated crests, $\nu$ could have been absorbed in the normalization of the pole density, since the flame slopes \eqref{slop2} or \eqref{sll} only depend on $\nu$ and $N$ through the combination $\mathcal{N} = \nu N$. This parameter $\mathcal{N}$ could then be \emph{any positive real number}, labeling a continuum of solutions. Without invoking pole decompositions, can one retrieve the selection of a discrete set of solutions (e.g. requiring $N = \mathcal{N}/\nu$ to be an integer) if small curvature effects ($0 < \nu \ll 1$) are restored afterward ? This was answered in the affirmative for viscous fingering or needle-crystals \cite{Pelce2}: quantization then resulted from a solvability condition on short-scale steady front-shape perturbations (local wavenumbers of order $1/\nu$). The corresponding procedure for flames has so far not been provided, and would be tantamount to performing directly a WKB analysis in \eqref{ZTW}.

A related matter concerns the solutions with integer $p > 1$ first encountered in \S~\ref{sec31} but discarded to ensure positivity of the pole density. We have shown in \S~\ref{sec22} that they nonetheless lead to solutions of the "inviscid ZT equation", i.e. \eqref{11} without the $\nu\phi_{xx}$ term. The $p > 1$ oscillating $\phi_x(x)$ profiles obtained for $0 < c < 1$ could then be viewed as steady nonlinear perturbations of the $p = 1$ solution \eqref{sll}, in a way reminiscent of what happens in needle-crystal growth \cite{Pelce2}. 

Besides, other equations than \eqref{11} may reduce to the inviscid ZT form in the large wrinkle limit, at least outside the cusps they might have. In case the same outer profiles as described by \eqref{varr}-\eqref{slop2} would apply, one would need to determine the appropriate $\mathcal{N}$ that enters in $B_{\max}$ in such equations as \eqref{bmax2}. One may guess that the inner front tip structure yields $\mathcal{N}$ (e.g., by matching), and one may wonder whether its value is quantized.

We also argued (the details are found in Appendix~\ref{appE}) that the ZT equation in the form \eqref{ZTW} appears in the $\beta$-deformation of the $\On$ random matrix model \eqref{qi}, in the limit of large matrices. One may wonder whether this very matrix model for finite $N$ (and its generalization where the eigenvalue lives in a certain region of the complex plane) could provide itself a statistical-mechanical description of a combustion problem.

\appendix
\section{The $c \rightarrow 1^{-}$ limit}
\label{appA}
The then strong coupling between $B_k$ of unlike signs in \eqref{22} makes the pole population split in two groups, separated by a distance of about $2h \geq \nu\,\frac{1 + c}{1 - c}$ that will exceed the width of each group for $c \rightarrow 1^{-}$. With $B_{j < 0} \approx -h$ to leading order, \eqref{23} written for $B_1,\ldots,B_N$ simplifies to:
\beq
\label{A1}\sum_{\substack{j = 1 \\ j \neq k}}^{N} \frac{2\nu}{(B_k + h) - (B_j + h)} + 2\,\frac{1 + c}{1 - c}\,\frac{\nu N}{B_k + h} \approx 1.
\eeq
According to Stieltjes (see \cite{Szego}, the review \cite{Marcellan} and references therein), the solutions $\xi_k = \frac{B_k + h}{\nu}$ to \eqref{A1} are the zeros of the associated Laguerre polynomial $\mathcal{L}_N^{(\alpha)}(\xi)$, with $\alpha = -1 + 2N\,\frac{1 + c}{1 - c}$. One may select $h$ so that $\sum_{j = 1}^N \frac{1}{B - (-B_j)} \approx \frac{N}{B + h}$ be exact at $B = h$, whereby $\frac{h}{\nu} = \frac{\alpha + 1}{2} = N\,\frac{1 + c}{1 - c}$ ; whatever $N$ is, $B_N - B_1 \propto \frac{\nu N}{\sqrt{1 - c}}$ \cite{IsmailLi} is asymptotically smaller than $h \sim \frac{2\nu N}{1 - c}$ when $c \rightarrow 1^{-}$, as anticipated. 
If next $N \rightarrow \infty$, the zeros of $\mathcal{L}_N^{(\alpha)}(\eta)$ are distributed according to a Mar\u{c}enko-Pastur density \cite{MarchenkoPastur}
\begin{equation}
\rho(\eta) \propto \frac{\sqrt{(\xi_{\max} - \xi)(\xi - \xi_{\min})}}{\xi},
\end{equation}
with $\xi_{\max/\min} = N(2 + a \pm 2\sqrt{1 + a})$ and $a = \lim_{N \rightarrow \infty} \frac{\alpha}{N} \approx 2\,\frac{1 + c}{1 - c}$. The ensuing pole density $\rho(B)$ is nonzero only for $B_{\min} < |B| < B_{\max}$, with $B_{\max(\min)} \approx \nu\xi_{\max(\min)} - h > 0$.

\section{Origin of the quantization \eqref{quanti}}
\label{appC}

We first recall the definition of the first Jacobi theta function:
\beq
\vartheta_1(w|\tau) = i \sum_{m \in \mathbb{Z}} (-1)^{p}\,e^{i\pi(2m - 1)w}\,e^{i\pi\tau(m + 1/2)^2}.
\eeq
This series is absolutely convergent whenever the parameter $\tau$ has positive imaginary part, it defines an entire function of $w \in \mathbb{C}$ with the following properties:
\bea
\vartheta_1(w + 1|\tau) & = & - \vartheta_1(\psi|\tau), \\
\qquad \vartheta_1(w + \tau|\tau) & = & - e^{-i\pi(2w + \tau)}\vartheta_1(w|\tau), \\
\vartheta_1(w = 0|\tau) & = & 0.
\eea
Therefore, we may build a function $\omega_+$ satisfying \eqref{omegaplus} with a ratio of theta functions. The choice of $\omega_+$ is arbitrary: indeed, the ratio of two functions satisfying \eqref{omegaplus} is $2K$- and $2{\rm i}K'$-periodic, so that it can be absorbed in the choice of $g(\psi)$ in \eqref{generalsolution}.
In this appendix, we take:
\beq
\label{imea}\omega_+(\psi) = \frac{\vartheta_1\big(\frac{\psi}{2{\rm i}K'} - {\rm i}\gamma/2|\tau)}{\vartheta_1(z|\tau)},\qquad \tau = \frac{{\rm i}K}{K'}.
\eeq
This function has a simple pole at $\psi = 0$ and a simple zero at $\psi = -\gamma K'$ (modulo the translations by $2K$ and $2{\rm i}K'$), unless:
\beq
\label{quan2}\gamma K' = 2Kp\qquad\mathrm{for}\,\,\mathrm{some}\,\,\mathrm{integer}\,\,p.
\eeq

We are now in position to explain why the case considered in \S~\ref{sec31} imposes the quantization condition \eqref{quan2}. Since the right-hand side in \eqref{34} was a constant, we were looking for a \emph{entire} function $\omega(\psi)$. The latter was decomposed as \eqref{generalsolution}, hence:
\beq
g(\psi)\omega_+(\psi) = \frac{e^{\pi\gamma}\omega(\psi + 2K) - \omega(\psi)}{e^{2\pi\gamma} - 1} - \frac{e^{\pi\gamma} - 1}{e^{2\pi\gamma} - 1}\,\frac{f}{2 + \nn}
\eeq
is also an entire function (we recall the assumption $c \neq 0$, thus $\gamma > 0$). When \eqref{quan2} is not satisfied, the choice of \eqref{imea} would imply that $g(\psi)$ has a simple pole at $\psi = -\gamma K'$, and no other singularity in $\mathbb{C}$. But no such function exists, because the total order of poles (modulo translations) of a meromorphic doubly-periodic function is at least $2$. The quantization \eqref{quan2} was thus necessary for our problem to admit solutions.

\section{$\On$ matrix models and ZT equation}
\label{appE}

Let us consider the statistical ensemble of $N$ particles at position $B_1,\ldots,B_N$ on the positive real axis, distributed according to the measure:
\beq
\label{qi}\prod_{i = 1}^{N} \dd B_i\,e^{-\beta \mathcal{V}(B_i)/2} \times \frac{\prod_{1 \leq i < j \leq N} |B_i - B_j|^{\beta}}{\prod_{1 \leq i,j \leq N} (B_i + B_j)^{\nn\beta/4}},
\eeq
where $\mathcal{V}(B)$ is a given, smooth function.  For $\beta = 2$, this is the $\On$ model introduced by \cite{GaudinKostov,Kostov} in relation with the problem of counting configurations of self-avoiding loops on random discrete surfaces. This model has been studied intensively since then \cite{KostovStaudacher,EynardKristjansen,EynardKristjansen2,TheseBorot}, and also appears in the context of quantum intrication \cite{BorotNadal}. When $\beta = 2$ and $\nn$ is an integer, \eqref{qi} is the measure induced on eigenvalues $B_1,\ldots,B_N$ of a random hermitian, positive definite matrix $\mathbf{B}$, and coupled to $\nn$ other hermitian matrices $\mathbf{A}_1,\ldots,\mathbf{A}_{\nn}$ with a joint distribution:
\beq
\label{Onem}\dd \mathbf{B}\,\prod_{j = 1}^{\nn} \dd \mathbf{A}_j\,e^{-\mathrm{Tr}[\mathcal{V}(\mathbf{B}) + \sum_{j = 1}^{\nn} \mathbf{B}\mathbf{A}_j^2]}.
\eeq
The quantity $\Delta(\mathbf{z})^2 = \prod_{i < j} |z_i - z_j|^{2}$ is characteristic of the eigenvalue distribution of random hermitian matrices whose full distribution is invariant under conjugation by a unitary matrix: it expresses the repulsion between eigenvalues of such a matrix taken at random. The product in the denominator of \eqref{qi} arises from the integration over the matrices $\mathbf{A}_j$ (which have a Gaussian distribution according to \eqref{Onem}).  In random matrix theory, \eqref{Onem} is called "the $\On$ matrix model", and replacing $\Delta(\mathbf{z})^2$ by $\Delta(\mathbf{z})^{\beta}$ is the "$\beta$-deformation".

In this model, the $B_k$ are random. We can define again a resolvent as an expectation value against the measure \eqref{qi}:
\beq
W(z) = \Big\langle \sum_{i = 1}^N \frac{1}{z - B_i} \Big\rangle,
\eeq
We claim that, when $N$ is large and with the choice $\mathcal{V}(B) = B/\nu$, $W(z)$ satisfies an equation like \eqref{ZTW}, which amounts to saying that $\phi_x(x) = \frac{4\nu}{1 - c}\,\mathrm{Im}[W({\rm i}x)]$ is solution of a ZT equation (see \S~\ref{sec22}).

It is convenient to introduce the two-points resolvent:
\beq
W(z_1,z_2) = \Big\langle \sum_{j,k = 1}^N \frac{1}{z_1 - B_j}\,\frac{1}{z_2 - B_k} \Big\rangle.
\eeq
In general $W(z_1,z_2) \neq W(z_1)W(z_2)$. However, when $N$ becomes large, the random $B_k$ for $k = 1,\ldots,N$ are distributed in a deterministic way, with some density $\varrho(B)$ supported on $\mathcal{C}$. Hence:
\beq
\label{conc}W(z_1,z_2) \sim W(z_1)W(z_2)\qquad\mathrm{when}\,\,N \rightarrow \infty.
\eeq
Using integration by parts, one can derive the Schwinger-Dyson relations valid for any $N$ \cite{TheseBorot}:
\bea
\label{resdu} & & W(z,z) + W(-z,-z) + \nn W(z,-z) \\
& & + \Big(\frac{2}{\beta} - 1\Big)\big[W'(z) + W'(-z)\big] - \frac{1}{\nu}\big[W(z) + W(-z)\big] = 0. \nonumber
\eea
This can be simplified when $N$ is large owing to \eqref{conc}:
\bea
& & W(z)^2 + W(-z)^2 + \nn W(z)W(-z) \\
& & + \Big(\frac{2}{\beta} - 1\Big)\big[W'(z) + W'(-z)\big] - \frac{1}{\nu}\big[W(z) + W(-z)\big] = 0. \nonumber
\eea
which coincides, up to the rescaling $z \rightarrow \frac{z}{2/\beta - 1}$, with \eqref{ZTW}.

\bibliography{Bibliflammes}

\end{document}